\documentclass[manuscript,screen,nonacm]{acmart}

\usepackage[english]{babel}

\usepackage{algorithm}
\usepackage{algpseudocode}
\usepackage{multirow}
\usepackage{xspace}
\usepackage{enumitem}
\usepackage{subcaption}
\usepackage{orcidlink}

\newcommand{\system}{\textsc{DynQ}\xspace}

\title{\system: A Dynamic Topology-Agnostic Quantum Virtual Machine via Quality-Weighted Community Detection}

\author{Shusen Liu\orcidlink{0000-0002-0794-1132}}
\affiliation{%
  \institution{Pawsey Supercomputing Research Centre}
  \city{Perth}
  \country{Australia}}
\affiliation{%
  \institution{The University of Western Australia}
  \city{Perth}
  \country{Australia}}

\author{Pascal Jahan Elahi\orcidlink{0000-0002-6154-7224}}
\authornote{Corresponding author.}
\email{pascal.elahi@csiro.au}
\affiliation{%
  \institution{Pawsey Supercomputing Research Centre}
  \city{Perth}
  \country{Australia}}
\affiliation{%
  \institution{The University of Western Australia}
  \city{Perth}
  \country{Australia}}

\author{Ugo Varetto\orcidlink{0000-0002-7696-0345}}
\affiliation{%
  \institution{Pawsey Supercomputing Research Centre}
  \city{Perth}
  \country{Australia}}
\affiliation{%
  \institution{The University of Western Australia}
  \city{Perth}
  \country{Australia}}

\begin{document}

\begin{abstract}

Quantum cloud platforms have scaled hardware capacity but not the abstraction exposed to users: small programs still monopolise entire processors, and existing Quantum Virtual Machine (QVM) designs often rely on fixed, topology-specific partitions that are brittle under calibration drift, spatial hardware heterogeneity, and transient defects. We present \system, a dynamic topology-agnostic QVM that derives virtual execution regions directly from live calibration data. \system models a processor as a quality-weighted coupling graph and formulates region discovery as community detection, thereby turning the classical systems principle of high internal cohesion and low external coupling into a hardware-aware objective for quantum virtualisation. This graph-driven formulation produces execution regions that are compilation-friendly, quality-aware, and resilient to degraded couplers and unavailable qubits.

\system further separates offline region discovery from online allocation, enabling low-latency scheduling over pre-validated regions while allowing the virtual layout to be recomputed whenever hardware conditions change.

Across calibration-derived simulations on five IBM backends, real-device experiments on IBM Kingston and Torino, and cross-architecture experiments on Rigetti Ankaa-3 via AWS Braket, \system improves the execution-quality floor, recovers workloads that the default mapping can lose under transient defects, and preserves stable output quality under concurrent batching. On the most heterogeneous simulated backend, \system reduces $L1$ error by up to 45.1\% and improves the output-similarity score by up to 19.1\%; on real devices, it eliminates the observed baseline failures in our workload. These results position quantum virtualisation as a graph-driven systems problem and show that adaptive, quality-aware QVMs can provide a practical substrate for reliable multi-tenant quantum cloud services.
\end{abstract}


\maketitle
\section{Introduction}

The classical cloud computing revolution was built on virtualisation, namely the ability to partition physical servers into isolated virtual machines serving multiple tenants simultaneously.
This fundamental abstraction transformed the economics of computing, enabling resource sharing, workload isolation, and differentiated quality of service (QoS). Despite its rapid growth, quantum cloud computing lacks this foundational capability. Platforms such as IBM Quantum~\cite{IBMQuantum2025a}, Amazon Braket~\cite{AWS2025a}, and Google Quantum AI~\cite{GQA2025} support users through a primitive sequential execution model: each program monopolises an entire quantum processor, regardless of size, precluding the multi-tenant resource sharing that defines modern cloud services.

IBM's 156-qubit Heron r2 and 133-qubit Heron r1 processors~\cite{IBMQuantum2025} routinely execute 5-qubit algorithms, leaving over 97\% of the hardware idle. With processors projected to reach thousands of qubits, this inefficiency will worsen dramatically. More critically, the sequential execution model prevents the economic scaling that classical clouds achieve through resource sharing; quantum cloud providers cannot amortise the cost of expensive hardware across concurrent users. Furthermore, on large devices, such sparse compilations exacerbate a well-known compiler pathology: conventional routing and layout heuristics operating on a huge coupling graph are prone to becoming trapped in poor local optima.

The concept of \emph{Quantum Virtual Machines} (QVMs) aims to bridge this gap by virtualising quantum resources, enabling isolation, sharing, and scalable utilisation of quantum processors. Broadly, existing QVM efforts fall into two conceptual categories. \emph{Hardware-level QVMs} partition a single quantum processor into multiple isolated execution regions to support multi-tenant concurrency, while \emph{program-level QVMs} virtualise execution through circuit fragmentation, gate virtualisation, or post-processing to extend the effective scale of computation beyond native hardware limits. Although these approaches address different bottlenecks, they share a common goal: decoupling logical quantum workloads from rigid physical hardware constraints.

Despite rapid progress in quantum hardware and cloud access models, existing QVM designs still fall short of providing a robust, general-purpose abstraction. Current approaches essentially treat virtualisation as a static mapping problem~\cite{Tao2025,tornow2024scaling}, binding virtual machines to fixed topological assumptions, nominal calibration snapshots, or coarse isolation boundaries. While such designs can be effective, they struggle to scale as devices grow larger, architectures diversify, and hardware behaviour becomes increasingly non-stationary. We elaborate on these limitations and position prior work in Section~\ref{sec:existing_qvm_limitations}.

\textbf{Our approach.} We present \system, a dynamic topology-agnostic QVM that derives execution regions from calibration-weighted device graphs. The central idea is to favour regions with high internal cohesion and low external coupling. Rather than imposing external geometric patterns, \system discovers natural QVM boundaries through quality-weighted community detection. \system models the quantum processor as a weighted graph in which edge weights encode gate fidelities, and applies the Louvain algorithm~\cite{Blondel2008}, a modularity-maximising community detector, to discover hardware regions with high internal quality and low boundary crosstalk. This approach requires no topology-specific template engineering: the same graph-based procedure can be applied across disparate coupling structures, including IBM's heavy-hex~\cite{Jurcevic2021}, Rigetti's square-lattice~\cite{RigettiComputing2025}, and, in principle, IQM-style star-connected architectures~\cite {Renger2025}.

The community detection formulation directly addresses the crosstalk-isolation challenge inherent in multi-tenant execution. In several quantum modalities (such as superconducting quantum hardware), couplers with low gate fidelity often exhibit elevated crosstalk, arising from frequency proximity, parasitic interactions, or strong residual coupling. By assigning lower weights to such edges, the modularity objective naturally places region boundaries along these low-quality couplers, separating tenants across the most interference-prone links. This results in implicit buffer zones that suppress cross-region error propagation without reserving dedicated idle qubits. Our dynamic re-discovery also adapts to transient defects by regenerating regions from current calibration data. 

Within each discovered community, standard quantum compilation techniques can be applied to optimise circuit-to-qubit placement and routing. \system\ therefore operates at a higher level than conventional compilers and can integrate with existing mapping algorithms such as SABRE~\cite{Zou2024} and noise-aware approaches, including TRAM~\cite{huang2025tram}.

Across calibration-derived simulations and real-device experiments on IBM backends, together with cross-architecture validation on Rigetti Ankaa-3, we show that \system raises the performance floor, recovers failed executions under unfavourable hardware conditions, and preserves output quality under concurrent batching.

\textbf{Contributions.} This paper makes the following contributions:

\begin{itemize}[leftmargin=*,nosep]
\item \textbf{A dynamic topology-agnostic QVM architecture.} We introduce a QVM design that derives virtual execution regions directly from calibration-weighted device graphs and decouples offline region discovery from online allocation.

\item \textbf{A quality-weighted partitioning formulation.} We formalise QVM region discovery as community detection on a quality-weighted coupling graph and introduce a region-scoring model that ranks candidate regions by connectivity, gate quality, readout quality, and internal uniformity.

\item \textbf{Robustness to transient hardware degradation.} We show that calibration-driven re-discovery allows \system to avoid degraded couplers and unavailable qubits, recovering executions that can collapse under static or default mappings.

\item \textbf{A multi-backend empirical evaluation.} We evaluate \system on five simulated IBM backends, two real IBM devices, and the Rigetti Ankaa-3 architecture, showing improved execution-quality floor, reduced worst-case failures, and stable batching behaviour under increasing concurrency.
\end{itemize}

\section{Background}

We briefly review the virtualisation principles that motivate our design, the architectural diversity of contemporary quantum processors, the noise mechanisms relevant to multi-tenant execution, and the limitations of existing QVM abstractions.

\subsection{From Classical to Quantum Virtual Machines}

Classical virtual machines (VMs) revolutionised computing infrastructure by enabling multiple isolated execution environments on shared physical hardware~\cite{popek1974formal}. The key principles underlying classical VM success translate directly to quantum virtualisation:

\paragraph{Resource partitioning:} Physical resources (CPU cores, memory, storage) are divided into isolated pools assigned to different tenants. Each VM sees a virtualised view of hardware resources, unaware of other tenants~\cite{rosenblum2005virtual}. In quantum systems, this corresponds to partitioning qubits and couplers into disjoint regions, each presenting an independent virtual quantum processor to its assigned program.
\paragraph{Strong isolation:} Classical hypervisors enforce isolation between virtual machines through hardware-supported memory protection: Memory Protection Keys~\cite{Park2019}; Capability Hardware Enhanced RISC Instructions~\cite{Watson2019}; Memory Integrity Enforcement~\cite{ASR2025}; privilege separation~\cite{Brumley2004}; and controlled resource multiplexing~\cite{Tan2019}. Resource contention or information leakage is prevented or strictly bounded by design. Quantum isolation is fundamentally more challenging. Beyond resource contention, quantum systems exhibit intrinsic non-classical correlations across logical separations: quantum states may become unintentionally entangled due to residual coupling, crosstalk, or imperfect control. These effects propagate coherently across the hardware substrate~\cite{Murali2020}. Consequently, effective quantum virtualisation requires isolation boundaries that are explicitly aligned with the underlying hardware physics to attenuate unintended quantum correlations between regions.
\paragraph{Quality-of-service (QoS) differentiation:} Classical cloud providers offer tiered service levels. For instance, AWS~\cite{AWS2025} offers on-demand, reserved, and spot instances, and Google Cloud Platform~\cite{GoogleCloud2025} offers Standard and Premium network service tiers. These services offer different performance and pricing trade-offs. Quantum QoS differentiation requires analogous quality stratification: identifying high-fidelity regions for premium tenants and lower-quality regions for best-effort workloads. Unlike classical workloads, some quantum applications, such as variational quantum algorithms (VQAs), exhibit noise tolerance and can operate in moderate-quality regions, whereas other algorithms require high-fidelity execution to produce useful results.
\paragraph{Dynamic adaptation:} Modern hypervisors and orchestration systems such as Kubernetes support dynamic resource management capabilities~\cite{Kubernetes2025} (e.g., live migration of virtual machines, horizontal and vertical auto-scaling of workloads, and automatic fault recovery). Of interest here is Kubernetes's automatic adjustment of the number of pods based on demand, and its rescheduling of failed workloads to maintain availability and performance. Quantum systems exhibit even more severe dynamic behaviour. In addition to transient defects that can disable qubits or couplers between calibration cycles, qubit parameters such as coherence times and gate fidelities drift over time due to environmental fluctuations and calibration adjustments. Any static region definition is brittle, and an effective QVM must therefore continually adapt region boundaries and allocations in response to evolving calibration data.

The related theoretical foundations of quantum programming and resource management have been established in prior work. Ying's foundational treatment~\cite{ying2016foundations} develops formal semantics for quantum programs, providing a rigorous basis for reasoning about quantum resource allocation. Recent work on quantum resource theories~\cite{chitambar2019} characterises quantum resources in terms of their inter-convertibility under allowed operations, though practical QVM systems must also address the physical constraints of real hardware.

\subsection{Quantum Processor Architecture}

Many contemporary quantum processors adopt fixed-topology architectures with sparse and irregular connectivity. Superconducting processors represent the most mature and widely deployed example of this class of systems. As such, we briefly summarise this technological modality. 

\begin{figure}[t]
  \centering
  \begin{subfigure}[b]{0.48\textwidth}
    \centering
     \Description{Connectivity graph of the IBM Heron r3 quantum processor showing qubits arranged in a sparse heavy-hex-like layout.}
    \includegraphics[width=\linewidth, keepaspectratio]{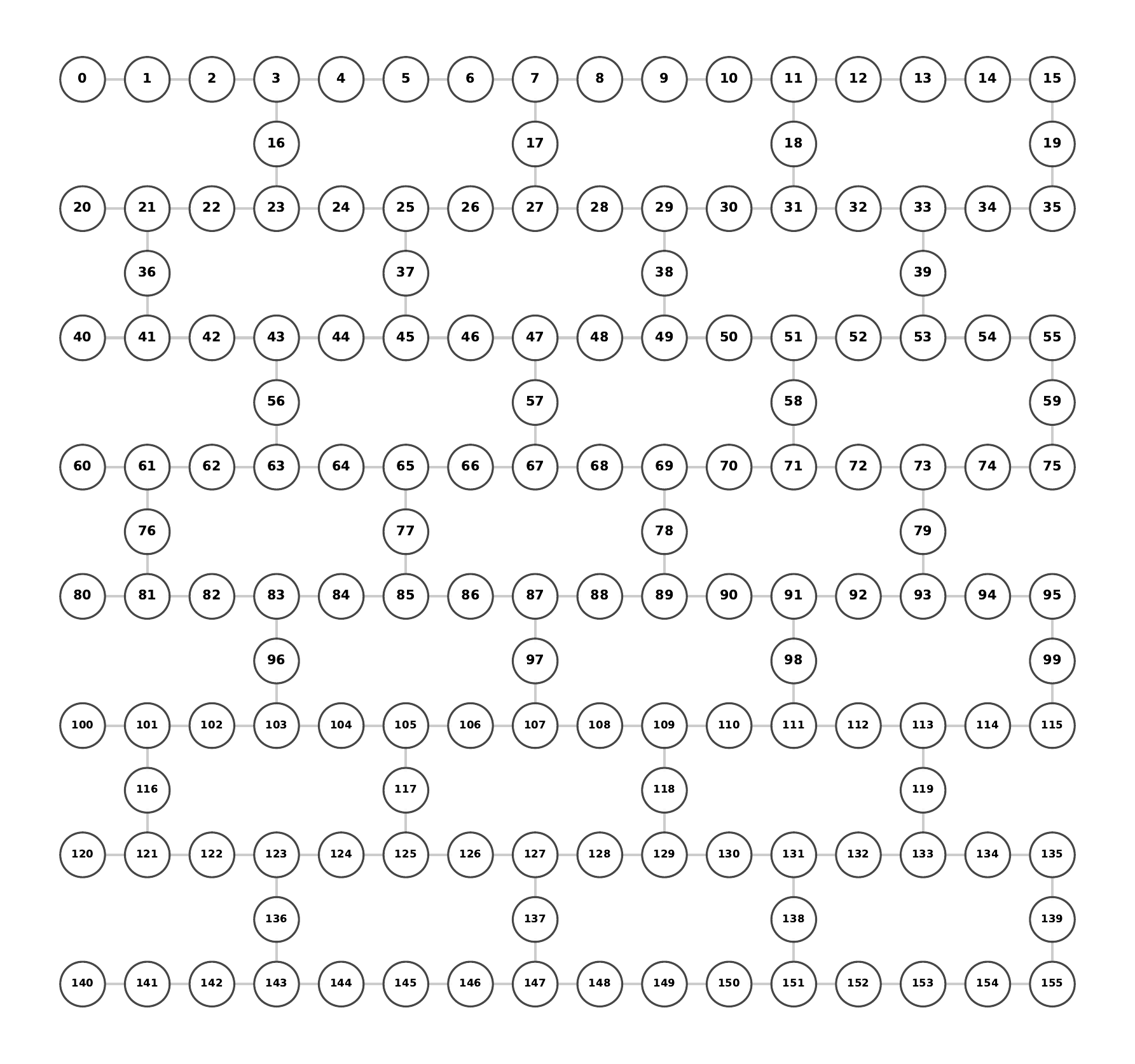}
    \caption{IBM Heron r3 series}
    \label{fig:ibm_heron}
  \end{subfigure}
  \hfill 
  \begin{subfigure}[b]{0.48\textwidth}
    \centering
    \Description{Connectivity graph of the Rigetti Ankaa-3 quantum processor showing a lattice-based qubit layout with local interactions.}
    \includegraphics[width=\linewidth, keepaspectratio]{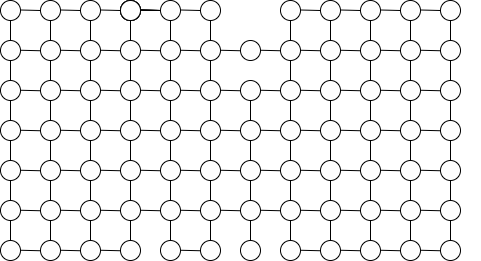}
    \caption{Rigetti Ankaa-3}
    \label{fig:ankaa}
  \end{subfigure}

\caption{\textbf{Representative coupling maps of superconducting quantum processors.}
Each node denotes a physical qubit, while edges indicate qubit pairs supporting native two-qubit gates.
The illustrated coupling graphs highlight the structural diversity and sparsity of contemporary superconducting hardware.
(a) IBM Heron r3 series adopts a heavy-hex-inspired lattice with deliberately reduced connectivity to mitigate frequency collisions and crosstalk.
(b) Rigetti Ankaa-3 follows a lattice-based layout with localised connectivity constraints.
Despite their differing geometric organisations, both architectures exhibit irregular, non-uniform coupling patterns, posing significant challenges for topology-aware quantum compilation and motivating the need for hardware-agnostic allocation and mapping strategies.}
    \label{fig:two_topologies}
\end{figure}

Superconducting quantum processors are composed of physical qubits, most commonly implemented as transmon qubits, which are nonlinear superconducting circuits designed to reduce sensitivity to charge noise. Each qubit interacts with others through engineered couplers such as resonators or capacitive connections, and the set of supported two-qubit interactions is described by the processor's \emph{coupling map}. Unlike classical processors with uniform interconnects, superconducting qubit coupling maps are inherently sparse and irregular due to constraints on physical layout and the need to avoid unwanted interactions and frequency crowding between neighbouring qubits. Figure~\ref{fig:two_topologies} provides representative examples of such fixed-topology quantum architectures. 

Limited connectivity is a deliberate design choice that enables individual qubit control and high-fidelity operations, but it also means that gates between non-adjacent qubits must be realised via sequences of SWAP operations or other routing primitives. Such routing incurs additional overhead and increases effective error rates as circuits grow in size and depth~\cite{Zou2024}. 

Physical qubits do not have identical physical characteristics, such as the coherence times $T_1$ (energy relaxation) and $T_2$ (dephasing), which quantify how long quantum information can be reliably preserved. On current superconducting platforms, these heterogeneous quality metrics are obtained through periodic calibration, typically performed daily or after thermal cycles.  Coherence times are typically on the order of 100--500~$\mu$s for IBM devices, 10--50~$\mu$s for Rigetti hardware~\cite{RigettiComputing2025}, and comparable ranges for contemporary IQM systems. These coherence limits impose a hard upper bound on the effective circuit depth: all gate operations must be completed before quantum states decohere.

In addition, gate errors that capture the probability of faulty unitary operations arising from imperfectly calibrated microwave pulses vary from qubit to qubit. Single-qubit gate error rates typically range from $10^{-4}$ to $10^{-3}$. Readout errors, typically on the order of $10^{-3}$ to $10^{-2}$, further contribute to overall execution infidelity due to imperfect discrimination between the $\lvert 0\rangle$ and $\lvert 1\rangle$ measurement outcomes.

Two-qubit gates are the dominant source of error in near-term quantum computation, as they require precise coordination between qubits and are inherently more sensitive to control noise, residual coupling, and calibration imperfections than single-qubit operations. For most superconducting platforms, the error rates of native two-qubit gates are typically $10^{-3}$ to $10^{-2}$, an order of magnitude higher than single-qubit gate error rates. Although the physical realisation of two-qubit gates varies across hardware platforms, their performance is consistently governed by local device parameters. \footnote{Fidelity depends on factors such as qubit frequency detuning, effective coupling strength, crosstalk from nearby control lines, and sensitivity to environmental noise. These parameters are shaped by fabrication variability and the local electromagnetic environment, leading to pronounced spatial heterogeneity in gate quality across the chip. For example, our experiments on a 156-qubit processor show error rates of $0.3\%$ to over $3\%$, with errors exhibiting strong spatial clustering rather than being uniformly distributed. }

This combination of fixed connectivity and heterogeneous qubit quality is common across several quantum hardware modalities and motivates topology-aware allocation and compilation strategies that jointly account for topology and qubit quality.

\subsection{Quantum Processor Topologies}

Different quantum hardware vendors have adopted distinct connectivity topologies, each with implications for virtualisation. Understanding these differences is essential for designing topology-agnostic QVM systems.

\paragraph{IBM Heavy-Hex Topology} IBM's Eagle (127 qubits), Heron (133 qubits and 156 qubits), and Condor (1121 qubits) processor families employ the heavy-hex topology~\cite{IBMQuantum2025}, a sparse superconducting coupling graph designed to support surface code and other nearest-neighbour quantum error correction schemes. In this layout, qubits are arranged in an augmented hexagonal pattern with additional “bridge” qubits that connect neighbouring hexagons. The resulting structure ensures that no qubit has more than three two-qubit neighbours, yielding an average degree of approximately three for a processor with $n$ qubits.

This constrained connectivity arises from the need to balance planar nearest-neighbour interactions with the need to avoid frequency collisions and excessive crosstalk. Compared to an idealised rectangular grid with degree-four connectivity, heavy-hex imposes irregular neighbourhoods and missing links, making many regular tilings or region shapes infeasible. As a result, multiprogramming and partitioning techniques that assume regular degree-4 grids or simple rectilinear regions cannot be directly applied on heavy-hex hardware, because such patterns either do not map cleanly to the available edges or leave qubits functionally isolated at region boundaries.

\paragraph{Rigetti Square-Lattice Topology} Rigetti's Ankaa-3 processor (84 qubits) adopts a square-lattice topology~\cite{RigettiComputing2025} in which interior qubits exhibit four-fold connectivity. This regular structure closely resembles an ideal two-dimensional grid, enabling straightforward rectangular tiling and simplifying circuit routing compared to more irregular coupling maps. In addition, Rigetti’s architecture incorporates tunable couplers, which allow the effective inter-qubit coupling strength to be dynamically adjusted and can be leveraged to mitigate crosstalk in ways not possible with fixed-coupling designs.

The regularity of the square lattice makes it a natural testbed for early multiprogramming and partitioning approaches, many of which implicitly assume grid-like connectivity when defining region shapes. However, geometric regularity alone does not imply uniform execution quality. Similar to IBM devices, Rigetti processors exhibit substantial spatial variation in two-qubit gate fidelity and crosstalk across the chip. As a result, even on square-lattice hardware, static geometric partitioning fails to capture quality heterogeneity, motivating quality-aware region formation irrespective of topology regularity.

\paragraph{IQM Square-Lattice Topology} IQM's current cloud-accessible processors, including the 20-qubit Garnet and 54-qubit Emerald, adopt a high-connectivity square-lattice topology with tunable couplers~\cite{IQC2025}. 
This design supports nearest-neighbour interactions with relatively dense connectivity and is compatible with surface-code error correction. 
The native gate set includes CZ gates with typical fidelities above 99\%.

From a QVM perspective, the lattice provides flexible but non-uniform connectivity, where logical regions must be discovered dynamically rather than relying on explicit structural boundaries. Notably, IQM has also proposed a star-resonator-based architecture enabling effective all-to-all connectivity via a central computational resonator. While not yet widely deployed in cloud-accessible systems, this design introduces natural hub-based interaction patterns that could fundamentally simplify region discovery in QVM abstractions.

\subsubsection{Implications for QVM Design.}
These topologies illustrate that QVM region definitions cannot be fixed templates. Even within a similar generation, like IBM Heron r1 and Heron r2, there can be significant differences in QVM structure and regional quality. A practical QVM layer must infer regions directly from the device graph so that it transfers across coupling structures without manual redesign. Moreover, connectivity alone is insufficient: because calibrated two-qubit fidelity and crosstalk vary spatially, regions should be formed that are internally high-quality, with boundaries placed around low-quality or high-interference links to enable QoS differentiation and robust isolation. Finally, the partition must be dynamic, since calibration updates and transient defects continuously reshape both the effective connectivity and the quality landscape.

\system satisfies these requirements with a single graph-based formulation. It constructs a quality-weighted coupling graph from calibration data and applies community detection to derive regions that adapt to topology, stratify by quality, and can be rediscovered as the device snapshot evolves.

\subsection{Error Propagation and Crosstalk in Multi-Tenant Execution}

Understanding error behaviour is essential for effective QVM partitioning, particularly in multi-tenant settings where multiple quantum programs execute concurrently on physically proximate hardware regions. Unlike classical computation, errors in quantum circuits do not remain locally confined. Imperfect operations introduce coherent and stochastic errors that propagate through entangling gates, accumulate with circuit depth, and, in shared hardware environments, can couple across nominal program boundaries. As a result, both intra-program error amplification and inter-program interference must be considered when designing isolation mechanisms for quantum virtual machines.

\paragraph{Intra-program error propagation} Within a single program, gate errors accumulate multiplicatively. For a circuit with $d$ two-qubit gates each having error rate $\epsilon$, the overall success probability is approximately $(1-\epsilon)^d \approx e^{-d\epsilon}$ for small $\epsilon$. This exponential decay explains why high-quality regions are critical: a region with half the error rate can support circuits twice as deep before reaching the same total error.

\paragraph{Inter-program crosstalk} More insidious in multi-tenant execution is crosstalk, namely interference between programs executing on nominally disjoint qubit regions. Even without shared qubits, adjacent regions influence each other through several mechanisms:

\emph{ZZ coupling} represents residual always-on interaction between neighbouring transmon qubits, arising from their physical capacitive coupling~\cite{Fors2024}. When two qubits are in states $|11\rangle$, they accumulate a relative phase $\phi_{ZZ} = 2\pi \xi_{ZZ} t$ where $\xi_{ZZ}$ is the $ZZ$ coupling strength (typically 10--100 kHz) and $t$ is the idle time~\cite{Fors2024}. This phase accumulation occurs even when no gates are applied, causing errors in both programs simultaneously. $ZZ$ coupling is strongest between directly-coupled qubits but extends to second neighbours through virtual photon exchange.

\emph{Measurement crosstalk} occurs when reading out one qubit disturbs the state of neighbours~\cite{Seo2021}. The readout pulse drives a resonator coupled to the target qubit; it can leak into adjacent resonators, leading to partial measurement of neighbouring qubits. On some devices, measuring qubit $i$ can flip the state of the neighbouring qubit $j$ with probability 0.1--1\%, corrupting the unmeasured qubit's quantum information.

\emph{Control line crosstalk} occurs when microwave pulses intended for one qubit partially affect others sharing the same control electronics or physically proximate control lines~\cite{Patterson2019}. A strong drive pulse on qubit $i$ may induce a weak (1--5\%) spurious rotation on qubit $j$~\cite{wang2022control}. This is particularly problematic when different tenants' programs apply gates simultaneously.

\paragraph{Crosstalk and QVM boundary placement} Effective QVM isolation requires boundaries that interrupt dominant error propagation paths rather than arbitrary geometric separations. In superconducting devices, the physical mechanisms responsible for inter-program crosstalk also degrade two-qubit gate fidelity, since both arise from strong residual coupling, frequency proximity, and shared control infrastructure. As a result, low-fidelity couplers identify links that are simultaneously computationally inefficient and prone to cross-region interference.

\system encodes this principle directly in its graph model. By weighting couplers according to calibrated two-qubit fidelity, community detection preferentially cuts along the weakest and noisiest links, yielding regions that are internally cohesive for high-quality execution while naturally attenuating inter-tenant interference at their boundaries.

\subsection{Transient Hardware Defects}

A critical but often underappreciated challenge for QVM systems is the prevalence of transient hardware defects. Unlike permanent fabrication defects, which can be identified and avoided offline, transient defects arise dynamically during device operation and can temporarily render otherwise functional qubits or couplers unreliable. These effects fundamentally undermine static assumptions about hardware availability and gate quality.

One major source of transient defects is calibration drift. Qubit frequencies and gate parameters evolve between calibration cycles due to slow environmental changes, charging effects, and material relaxation. As a result, two-qubit gates that are initially calibrated to high fidelity can degrade significantly within a single calibration window or fail entirely if frequency drift moves the interaction off-resonance. Although periodic calibration resets these parameters, defects may emerge unpredictably mid-cycle, leaving the effective device graph inconsistent with its nominal specification. Additional transient behaviour arises from microscopic two-level system fluctuators present in substrates and Josephson junction barriers. These parasitic quantum systems can intermittently couple to nearby qubits when their frequencies drift into resonance, causing abrupt reductions in coherence times. Such events are stochastic in nature, can persist from minutes to days, and may reduce qubit lifetimes by an order of magnitude without warning. Environmental perturbations further exacerbate this variability, including correlated quasiparticle bursts induced by cosmic rays or natural radioactivity, as well as spatially non-uniform thermal fluctuations, which can transiently degrade coherence and gate performance across large regions of the chip~\cite{Klimov2018,Mueller2019}.

For QVM systems, the implications of these transient defects are severe. Static partitioning approaches, such as template-based region definitions, assume stable connectivity and uniform availability within each region. In practice, a single degraded coupler or qubit can render an entire predefined region invalid, leaving the compiler unable to route around the defect. In our experiments, such events result in complete execution failures and zero observed fidelity when circuits traverse affected regions. Across real devices, we observe failure rates consistent with transiently impaired regions affecting approximately 5--10\% of static allocations at any given time.

\system addresses this challenge by treating hardware state as dynamic rather than static. Regions are derived from the current calibration snapshot, with disabled qubits removed from the graph and degraded couplers assigned low or zero weight. Community detection then naturally routes regions around transient defects instead of through them. When calibration data updates, re-running region discovery requires only 0.81\,s, producing an updated set of QVM regions that reflect the current hardware conditions and restore availability without manual intervention.

\subsection{Quantum Cloud Cost Models and Multiprogramming}

Quantum cloud pricing models drive economic incentives for multiprogramming. On current platforms such as IBM Quantum and AWS Braket, users are charged based on \emph{job execution time} and shots, with each job executing a circuit or a batch of circuits for a fixed number of shots, typically in the range of 1024--4096.  Importantly, the cost of a job is largely independent of the number of qubits used by the circuit. As a result, a small circuit occupying a few qubits incurs the same cost as a much larger circuit scheduled on the same backend.

This pricing structure creates a strong incentive for multi-tenant execution at the cloud level. By serving multiple independent user workloads within a single job submission, the provider can amortise the cost of quantum hardware across tenants, improving utilisation and throughput while reducing queueing delays. From the provider’s perspective, multiprogramming transforms quantum processors from single-tenant, batch-oriented resources into shared infrastructure more closely resembling classical cloud systems.

At the system level, \system assigns each circuit to a disjoint physical region of the processor. For circuits $C_1, \ldots, C_k$ requiring $n_1, \ldots, n_k$ logical qubits, a valid assignment maps each $C_i$ to a region $R_i$ such that $|R_i| \geq n_i$, the subgraph induced by $R_i$ is connected, and regions are pairwise disjoint to ensure isolation. Connectivity is essential because physical paths within each region must support logical two-qubit interactions. Sparse or poorly connected regions incur additional routing overhead through SWAP operations, increasing circuit depth and error accumulation. Consequently, effective multiprogramming using a QVM requires not only sufficient qubit count but also high-quality internal connectivity within each assigned region.

Ideally, this multi-tenancy should be exposed to users through abstraction rather than manual batching. Users typically submit individual circuits or workflows and expect the system to schedule them efficiently without requiring explicit coordination with other tenants. Quantum multiprogramming, implemented through a QVM layer, provides this abstraction by automatically combining independent circuits into a shared execution while preserving isolation and execution fidelity.

Consider a workload consisting of $N$ independent circuits. The execution cost differs fundamentally under the two execution regimes:
\begin{itemize}[nosep,leftmargin=*]
\item \textbf{Sequential execution:} each circuit is submitted as an individual job, resulting in $N$ job submissions and a total cost proportional to $N$.
\item \textbf{Batched execution via QVM multiprogramming:} up to $k$ circuits are combined into a single job, yielding approximately $\lceil N/k \rceil$ job submissions and a total cost proportional to $N/k$.
\end{itemize}
Under this simplified per-job accounting model, batching reduces the required number of job submissions by approximately a factor of $k$. This cost reduction is evaluated empirically in Section~\ref{sec:service}.

\subsection{Limitations of Existing QVM Approaches}
\label{sec:existing_qvm_limitations}

Prior work uses the term \emph{quantum virtual machine} to describe several mechanisms that improve the efficiency of quantum clouds. One major direction is spatial multiplexing (quantum multi-programming), where multiple independent circuits are mapped to disjoint chip subsets and executed concurrently to increase utilisation and reduce queueing delay~\cite{das2019case, liu2021qucloud, liu2024qucloudplus, Niu2023enabling}. A complementary direction introduces provider-side virtual regions as a runtime abstraction, allowing tenants to compile independently while the cloud assigns jobs to predefined areas at runtime~\cite {Tao2025}. A third, largely orthogonal, body of work explores gate virtualisation and circuit fragmentation to scale a single computation beyond device size or connectivity limits~\cite{tornow2024scaling}. Together, these lines demonstrate the promise of virtualisation, but they also expose structural gaps when viewed through the lens of a practical, multi-tenant cloud layer.

\paragraph{Batch-coupled placement and compilation.}
A recurring theme across multiprogramming systems is the tight coupling among batching, compilation, and placement. Many proposals implicitly assume that the set of co-running programs is known at mapping time, because partitioning and routing decisions are optimised for a specific batch composition~\cite{das2019case, liu2021qucloud, Niu2023enabling}. While reasonable for offline evaluations, this assumption is restrictive in cloud settings where workloads arrive asynchronously and the scheduler must balance latency, throughput, and fairness. As a result, the abstraction remains batch-specific rather than a stable, reusable, and consistently managed virtual resource across workload instances.

\paragraph{Topology-bound virtualisation and limited portability.}
Several QVM designs achieve high performance by exploiting detailed knowledge of the coupling graph through hand-crafted region templates or topology-tailored heuristics~\cite{liu2021qucloud, Tao2025}. HyperQ, for example, formalises the virtual-region abstraction and demonstrates efficient tiling on a target topology~\cite{Tao2025}. However, such designs do not generalise seamlessly as processor architectures diversify: region definitions and placement heuristics must be redesigned or retuned for each new coupling structure, creating an increasing engineering burden and limiting portability across hardware generations~\cite{liu2024qucloudplus}.

\paragraph{Brittleness under dynamic calibration and transient defects.}
Cloud-ready virtualisation must remain robust as the effective device graph evolves across calibration cycles and within calibration windows. In practice, qubits or couplers can degrade or become unavailable due to drift and transient defects, reshaping feasible connectivity without warning. Static region definitions are brittle under these dynamics: a single degraded link can invalidate an entire region, leading to allocation failures or excessive rerouting~\cite{Tao2025}. Compiler-based multi-programming often incorporates calibration data into one-shot mappings, but typically treats re-partitioning as an offline or exceptional process rather than a routine runtime operation, leaving a gap between hardware reality and the assumptions embedded in current abstractions~\cite{liu2021qucloud, Niu2023enabling}.

\paragraph{Quality-agnostic resource models and coarse isolation.}
Many approaches ensure feasibility by enforcing connectivity and disjointness, and some incorporate crosstalk-aware heuristics, yet they largely treat feasible regions as interchangeable~\cite{liu2024qucloudplus, Niu2023enabling}. On real devices, two-qubit gate fidelities and interference profiles vary substantially across qubit pairs. Provider-side isolation mechanisms such as reserving fixed buffer qubits between regions~\cite{Tao2025} can suppress interference, but they are inherently static: they do not adapt to spatial quality variations or transient calibration changes, and they may waste high-fidelity qubits by dedicating them to separation rather than computation. The absence of an explicit, quality-stratified resource model limits existing QVMs' ability to provide differentiated service levels and suppress inter-tenant interference as concurrency increases.

\paragraph{Scale-up virtualisation is orthogonal to multi-tenancy.}
Gate virtualisation techniques address the challenge of executing a single computation beyond hardware limits by transforming and distributing interactions~\cite{Tao2025}. While powerful, these methods target a different axis of scalability and do not directly address concurrent isolation, dynamic reconfiguration from calibration snapshots, or quality-aware placement within a shared device.

Taken together, existing QVM approaches illustrate the benefits of virtualisation but stop short of providing a unified cloud abstraction that is portable across topologies, resilient to calibration dynamics, and explicit about hardware quality. These gaps motivate a QVM design that treats the calibrated device graph as runtime state, derives virtual regions algorithmically rather than from fixed templates, and decouples tenant submission from placement decisions.

\section{\system Design}
\label{sec:design}

\system instantiates the QVM abstraction through a principled two-phase architecture: \emph{offline discovery} transforms hardware calibration data into a pool of quality-stratified virtual machines, while \emph{online allocation} maps tenant programs to available QVMs at runtime. This separation mirrors classical virtualisation: the hypervisor manages physical resources (offline) while presenting virtualised interfaces to guests (online).

The key design principle is \textbf{high cohesion, low coupling}, borrowed from classical systems design. Within each QVM region, we maximise internal connectivity (cohesion) to enable efficient circuit compilation with minimal SWAP overhead. Between regions, we minimise connection strength (coupling) to suppress crosstalk interference between tenants. The Louvain community detection algorithm~\cite{Blondel2008} naturally optimises this objective by maximising modularity.

We first describe the layered system architecture, then present the offline discovery algorithms of region selection, runtime allocation, and region-restricted compilation.

\subsection{System Architecture Overview}

\begin{figure}[!htp]
  \centering  
  \Description{Layered system architecture diagram showing separation between an offline topology discovery component and an online allocation component. A hardware graph abstraction is used to identify and store validated quantum virtual machine regions that are later assigned to incoming quantum programs.}
  \includegraphics[width=0.7\linewidth]{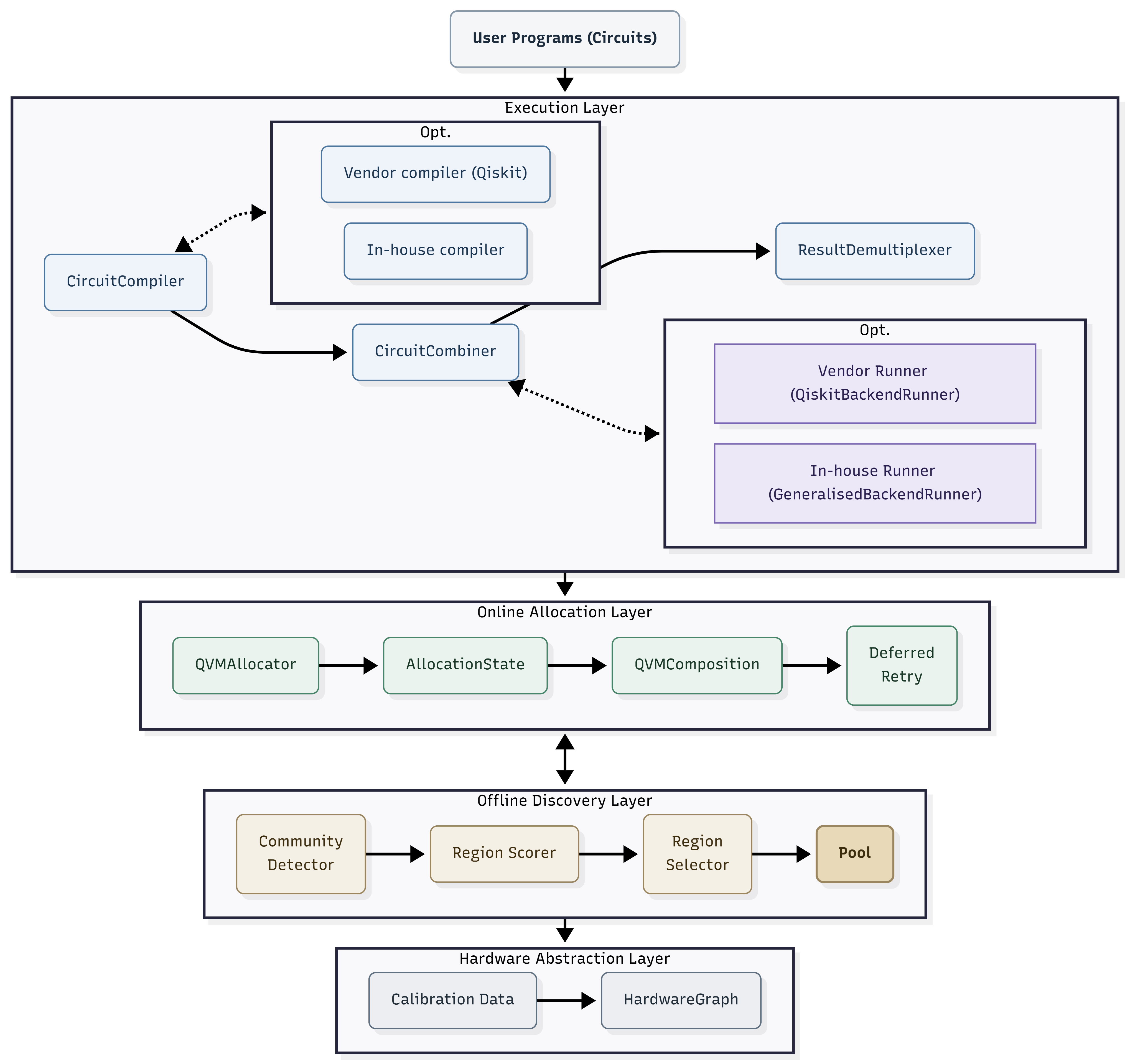}
\caption{\textbf{The layered architecture of \system.} The design explicitly decouples computationally intensive topology analysis (\textit{Offline Discovery Layer}) from latency-critical scheduling decisions (\textit{Online Allocation Layer}). By abstracting physical device properties into a \texttt{HardwareGraph}, the system maintains a pool of pre-validated QVM atomic regions to service incoming user programs efficiently.}
  \label{fig:architecture}
\end{figure}

Figure~\ref{fig:architecture} presents \system's layered architecture that separates topology- and quality-dependent analysis from latency-critical runtime decisions. This separation adapts to evolving hardware conditions while providing fast, predictable allocation for incoming workloads. At a high level, QVM functionality is decomposed into four layers of abstraction and decision-making.

At the lowest level, the \texttt{HardwareGraph} module serves as the \emph{hardware abstraction layer}. It constructs a uniform representation of the underlying quantum processor and extracts calibrated qubit- and coupler-level properties from the backend calibration interface. These properties are encoded in a weighted graph, with vertices representing qubits, edges representing native couplers, edge weights representing two-qubit gate quality, and vertex attributes storing per-qubit properties such as readout error.

Above this abstraction sits the \emph{offline region discovery stage}. Here, the \texttt{CommunityDetector} module applies modularity-based community detection to the weighted hardware graph, producing a set of densely connected subgraphs that serve as candidate QVM atomic regions. These candidate regions are then evaluated by the \texttt{RegionScorer} using multiple quality indicators, including internal gate fidelity and boundary weakness. Finally, the \texttt{RegionSelector} resolves overlaps and selects a non-overlapping subset, yielding a pool of pre-validated QVM regions annotated with quality scores. Because this computationally intensive analysis process runs offline per calibration cycle, it does not affect runtime latency and the cost is amortised over each calibration cycle.

The \emph{online allocation layer} handles real-time scheduling decisions for incoming circuits. The \texttt{QVMAllocator} selects a suitable precomputed region for each incoming circuit based on size, quality, and availability. The \texttt{AllocationState} maintains region occupancy to preserve isolation between concurrent tenants. Extensions for fragmented-capacity handling and deferred admission are discussed in Appendix~\ref{app:alloc-extensions}. When no suitable region is immediately available, the \texttt{DeferredRetryScheduler} defers the circuit to a subsequent batch, enabling graceful handling of contention without blocking the system. This design ensures that runtime allocation remains lightweight while still exploiting the richer structure discovered offline.

Finally, the \emph{execution layer} is responsible for assembling and running multi-tenant workloads. The \texttt{CircuitCompiler} forwards each circuit to a standard compilation backend, such as Qiskit’s default \texttt{qiskit.transpiler}, together with the topology and calibration metadata of the assigned QVM region. The compiler then performs region-restricted transpilation, and the resulting compiled circuit is collected and returned to the execution layer. The \texttt{CircuitCombiner} merges the circuits assigned to distinct QVM regions into a single composite circuit that preserves logical isolation. The resulting composite circuit is then remapped and passed to a backend runner abstraction for execution. This abstraction may be realised by the \texttt{QiskitBackendRunner}, which submits the circuit to a conventional backend, or by the \texttt{GeneralisedBackendRunner}, which exposes a unified remote invocation interface to one or more quantum processing units and can support more efficient execution paths. After execution, the \texttt{ResultDemultiplexer} separates measurement outcomes and transparently returns them to individual tenants via a reverse-remapping process, yielding the results for each tenant. From the user’s perspective, each circuit executes independently, while the system internally realises the efficiency gains of concurrent execution.

\subsection{Hardware Graph Construction}
We model the quantum processor as a weighted undirected graph $G = (V, E, w)$. Vertices $V$ represent physical qubits, edges $E$ represent couplers that enable two-qubit gates, and the weight function $w: E \rightarrow \mathbb{R}^+$ encodes gate quality.

We define edge weights as the inverse of gate error rates:
\begin{equation}
w_{ij} = \frac{1}{\epsilon_{ij} + \epsilon_0}
\end{equation}
where $\epsilon_{ij}$ is the two-qubit gate error rate for the coupler between qubits $i$ and $j$, and $\epsilon_0 = 10^{-6}$ is a small regularisation constant to avoid numerical instability. Here, $\epsilon_{ij}$ can be instantiated as the \emph{RB-derived average two-qubit gate infidelity} reported by the backend calibration interface\footnote{The formulation can be naturally extended to incorporate directional gate asymmetry (where $\epsilon_{i\rightarrow j} \neq \epsilon_{j\rightarrow i}$) and crosstalk penalties when such calibration data is available.}. This inverse relationship ensures that high-fidelity gates receive high weights. As community detection depends only on relative edge weights, any monotonic transformation of gate error rates yields equivalent partitions.

This weighting scheme has a natural interpretation in terms of circuit fidelity. For a circuit using $k$ two-qubit gates, the overall success probability is approximately $\prod_{i=1}^{k}(1-\epsilon_i) \approx e^{-\sum_i \epsilon_i}$ for small error rates, although two-qubit errors are not strictly independent Bernoulli events\cite{Sarovar2020detectingcrosstalk}. Minimising the total error contribution is equivalent to minimising $\sum_i \epsilon_i$, which our weighting encourages by favouring high-weight (low-error) edges.

The weighted graph naturally handles hardware defects without special-case logic. Disabled qubits are simply absent from the vertex set $V$; broken couplers are absent from the edge set $E$. When community detection is applied to this graph, it observes only functional hardware and partitions accordingly. Defects automatically become region boundaries because no edges cross them. 

Graph construction requires a single pass over calibration data and runs in $O(|V| + |E|)$ time. On a 156-qubit heavy-hex processor~\cite{IBMQuantum2025}, this overhead is negligible compared to the calibration interval.

\subsection{Community Detection: Formalising High Cohesion, Low Coupling}
Effective QVM partitioning can be cast as a graph-structural objective. Within a QVM atomic region, qubits should be connected via high-quality couplers to support efficient compilation while minimising SWAP overhead. In contrast, connections across regions should be weak to attenuate crosstalk and residual interactions between concurrent programs. In graph-theoretic terms, this corresponds to partitions with strong intra-regional connectivity and weak inter-regional connectivity.

Community detection, which is designed to identify densely connected subgraphs that are only sparsely linked to one another, directly captures the isolation and locality requirements of multi-tenant quantum execution. This formulation aligns with a long-standing principle in classical systems design, where modular structures are preferred for their strong internal coherence and limited external interaction.

We use the Louvain algorithm~\cite{Blondel2008} to optimise modularity. For a partition of graph $G$ into communities, modularity is defined as:
\begin{equation}
Q = \frac{1}{2W} \sum_{i,j} \left[ w_{ij} - \frac{s_i s_j}{2W} \right] \delta(c_i, c_j)
\end{equation}
where $w_{ij}$ is the edge weight between vertices $v_i$ and $v_j$ (zero if no edge exists), $s_i = \sum_j w_{ij}$ is the weighted degree of vertex $v_i$, $W = \frac{1}{2}\sum_{i,j} w_{ij}$ is the total edge weight in the graph, and $\delta(c_i, c_j)$ equals 1 if vertices $i$ and $j$ belong to the same community and 0 otherwise. Here, $Q$ is a scalar objective that measures partition quality, with larger values indicating stronger community structure. The interpretation is straightforward: modularity measures the partition's excess internal connectivity, given by the observed within-community weight $w_{ij}$, relative to the null-model expectation $\frac{s_i s_j}{2W}$ under a random graph that preserves the weighted degree sequence. 

Maximising modularity preferentially groups qubits connected by high-weight edges, corresponding to low-error couplers, into the same community, whereas low-weight edges contribute little to the objective and are more likely to span community boundaries. In effect, the optimisation encourages regions whose internal connectivity is supported by high-fidelity gates, while naturally placing partition boundaries along noisy or unreliable couplers. This behaviour is not imposed heuristically but emerges directly from the structure of the modularity function.

The Louvain algorithm optimises modularity through a greedy agglomerative process with two alternating phases, which halts when modularity no longer improves.
\paragraph{Local optimisation} The algorithm iteratively refines an initial partition by relocating individual vertices to neighbouring communities whenever doing so yields a positive modularity gain. This phase rapidly increases modularity by exploiting local graph structure and converges when no further-improving moves remain. In the context of \system’s quality-weighted hardware graph, this process preferentially groups qubits connected by high-fidelity couplers, since such moves yield the largest immediate modularity gains.
\paragraph{Aggregation} Once local optimisation stabilises, communities are contracted into super-vertices to form a coarsened graph, with edge weights aggregated accordingly. Modularity optimisation is then repeated on this reduced graph, enabling the algorithm to capture higher-level community structure beyond local optima. For \system, this hierarchical aggregation produces execution regions that remain cohesive across scales, allowing region boundaries to stabilise around consistently weak or noisy couplers rather than transient local fluctuations. 

This strategy balances solution quality with computational efficiency. Local moves exploit fine-grained structure, while aggregation enables global reorganisation at progressively coarser scales. In practice, the algorithm converges quickly, making it suitable for execution once per calibration cycle. For example, this process completes in $\lesssim100$~ms on a 156-qubit heavy-hex graph.

\subsection{Region Scoring}
\label{sec:scoring}

Not all discovered communities are equally suitable as QVM atomic regions. Community size directly affects both practical usability and scheduling flexibility in a multi-tenant setting. Very small communities, while potentially useful for trivial circuits or diagnostics, offer limited capacity and tend to increase fragmentation in the region pool. Conversely, very large communities may span heterogeneous quality zones, mixing high- and low-fidelity couplers in ways that undermine quality-aware placement. To balance flexibility, usability, and robustness, we prioritise communities with at least three qubits as primary QVM atomic regions and evaluate them using multiple complementary quality metrics that capture distinct aspects of region suitability. Also, to preserve full hardware coverage, residual single or two-qubit fragments may also be retained in the final partition.

For a candidate region $R$ (a connected subgraph with vertex set $V_R$ and edge set $E_R$), we compute four scores:

\paragraph {Connectivity score} Measures how densely connected the region is internally:
\begin{equation}
S_\text{conn}(R) = \frac{2|E_R|}{|V_R|(|V_R|-1)}.
\end{equation}
This is the ratio of actual edges to the maximum possible edges in a complete graph. For a fully-connected region, $S_\text{conn} = 1$; for a linear chain of $n$ qubits, $S_\text{conn} = 2/(n-1)$. Higher connectivity means more routing options during circuit compilation, reducing SWAP overhead. On heavy-hex topologies, typical regions achieve $S_\text{conn}$ between 0.2 and 0.5.

\paragraph{Gate quality score} Captures the average two-qubit gate fidelity:
\begin{equation}
\label{eq:gate_quality}
S_\text{gate}(R) = \max\left(0, 1 - \alpha_\text{gate} \bar{\epsilon}_\text{gate}\right), 
\end{equation}
where $\bar{\epsilon}_\text{gate} = \frac{1}{|E_R|}\sum_{(i,j) \in E_R} \epsilon_{ij}$ is the mean gate error rate across all couplers in $R$. The scaling factor of $\alpha_\text{gate}\sim100$ maps typical error rates to a 0--1 score: 0.5\% error yields $S_\text{gate} = 0.5$, while 1.5\% error yields a near-zero score. This score directly reflects the dominant source of error in NISQ circuits.

\paragraph{Readout quality score} Captures measurement fidelity:
\begin{equation}
S_\text{ro}(R) = \max\left(0, 1 - \alpha_\text{ro} \bar{\epsilon}_\text{ro}\right),
\end{equation}
where $\bar{\epsilon}_\text{ro} = \frac{1}{|V_R|}\sum_{i \in V_R} \epsilon_{\text{ro},i}$ is the mean readout error across qubits in $R$. Here we use $\alpha_\text{ro}=10$, which differs from gate quality because readout errors are typically larger (1--5\% versus 0.3--3\% for gates). Readout fidelity affects the final measurement step of every circuit and cannot be mitigated through circuit optimisation.

\paragraph{Uniformity score}
Penalises internal variation in two-qubit gate quality:
\begin{equation}
S_\text{unif}(R) = \max\left(0, 1 - \mathrm{CV}_\text{gate}(R)\right),
\end{equation}
where
\begin{equation}
\mathrm{CV}_\text{gate}(R) = \frac{\sigma_\epsilon}{\mu_\epsilon},
\end{equation}
is the coefficient of variation of two-qubit gate error rates within region $R$. Here, $\mu_\epsilon = \frac{1}{|E_R|}\sum_{(i,j)\in E_R}\epsilon_{ij}$ and 
$\sigma_\epsilon = \sqrt{\frac{1}{|E_R|}\sum_{(i,j)\in E_R}(\epsilon_{ij}-\mu_\epsilon)^2}$ 
denote the mean and standard deviation of two-qubit gate error rates within region $R$, respectively. High region uniformity allows compilers to apply consistent optimisation strategies. By contrast, a region with acceptable average quality but large internal variability may limit a compiler’s ability to exploit its strongest interactions without also traversing substantially weaker ones.

The composite quality score combines these metrics as a weighted sum:
\begin{equation}
Q(R) = S_{\text{conn}}(R) + \beta_\text{gate} S_{\text{gate}}(R) + \beta_\text{ro} S_{\text{ro}}(R) + \beta_\text{unif} S_{\text{unif}}(R).
\end{equation}
Gate quality is assigned the highest relative importance, reflecting that two-qubit gate errors typically dominate overall circuit fidelity on current NISQ hardware. Connectivity is also an important factor, as limited connectivity increases SWAP overhead and circuit depth during compilation. Readout fidelity and community uniformity are treated as secondary factors that help refine region quality but usually have a smaller impact on overall circuit fidelity.

The coefficients $\beta_\text{gate}$, $\beta_\text{ro}$, and $\beta_\text{unif}$ therefore control the relative contribution of these factors. In practice, the optimisation is not highly sensitive to the precise parameter values, similar to observations reported for TRAM~\cite{huang2025tram}. The coefficient of $S_{conn}$ is fixed to 1 and serves as a baseline. The remaining parameters are expressed relative to this term.
Because the objective is used only to rank candidate regions, the coefficients need not sum to 1. In our experiments, we use $\beta_\text{gate} = 1$ and $\beta_\text{ro} = \beta_\text{unif} = 0.5$, which provides a good balance between gate reliability and secondary quality metrics.

\subsection{QVM Atomic Region Selection}

Community detection typically yields a QVM atomic region \emph{pool} that contains
(i) overlapping candidates arising from the hierarchical refinement of partitions and
(ii) more candidates than can be instantiated simultaneously on the same processor.
The scheduling layer, therefore, requires a \emph{non-overlapping} subset of QVM atomic regions that achieves high aggregate quality while preserving sufficient flexibility for subsequent allocation.

Formally, let $\{R_i\}_{i=1}^m$ denote the set of candidate QVM atomic regions, where each region $R_i$ is associated with a quality score $Q(R_i)$ and occupies a set of physical qubits
$V(R_i) \subseteq V$. The region selection problem is defined as:
\begin{equation}
\begin{aligned}
\max_{\mathcal{S} \subseteq \{R_1,\dots,R_m\}} \quad & \sum_{R \in \mathcal{S}} Q(R) \\
\text{s.t.} \quad & V(R_i) \cap V(R_j) = \emptyset, \quad \forall R_i \neq R_j \in \mathcal{S}.
\end{aligned}
\end{equation}
This formulation corresponds to the weighted set packing problem and is NP-hard. Algorithm~\ref{alg:selection} outlines the approach: we sort candidates by quality density (score per qubit) and then greedily select non-conflicting regions.

\begin{algorithm}[t]
\caption{Greedy Region Selection}
\label{alg:selection}
\begin{algorithmic}[1]
\small
\Require Candidate regions $\{R_1, \ldots, R_m\}$ with scores $Q(R_i)$
\Ensure Selected regions $\mathcal{S}$
\State Sort candidates by $Q(R_i)/|R_i|$ in descending order
\State $\mathcal{S} \gets \emptyset$; $\text{used} \gets \emptyset$
\For{each $R$ in sorted order}
    \If{$R \cap \text{used} = \emptyset$}
        \State $\mathcal{S} \gets \mathcal{S} \cup \{R\}$
        \State $\text{used} \gets \text{used} \cup R$
    \EndIf
\EndFor
\State \Return $\mathcal{S}$
\end{algorithmic}
\end{algorithm}
Ranking by density biases selection toward compact high-quality regions and avoids a common failure mode in which a large but mediocre region blocks multiple smaller, higher-quality alternatives. To illustrate this, consider two competing selections that overlap in physical qubits. The first consists of a single region $R_A$ with $|V(R_A)| = 20$ and quality score $Q(R_A)=0.6$, yielding a density of $0.03$. The second consists of two disjoint regions $R_B$ and $R_C$, each with $|V|=10$ and $Q=0.5$, yielding densities of $0.05$. Although $R_A$ has higher absolute quality than either $R_B$ or $R_C$ individually, selecting $R_A$ excludes both smaller regions. In contrast, selecting $R_B$ and $R_C$ achieves a higher total utility
($1.0$ versus $0.6$) while preserving greater flexibility for subsequent allocation. Density-based ordering correctly prefers the latter outcome.

\subsection{Online Allocation}
\label{sec:online}

At runtime, \system maintains a pool of available QVM atomic regions (those not currently hosting executing programs) and services incoming allocation requests. The allocation policy prioritises close size matching, high region quality, and low routing overhead when selecting from the currently available regional pool.

For a request requiring $n$ qubits, \system evaluates each available region $R$ using the fitness function
\begin{equation}
F(R, n) = S_\text{size}(R,n)\bigl[\gamma_\text{conn}S_\text{conn}(R) + \gamma_Q Q(R)\bigr].
\end{equation}
We explicitly retain $S_\text{conn}$ as a separate term, even though it is included in $Q$, to bias runtime allocation toward regions that minimise routing overhead for the target circuit. In practice, $\gamma_\text{conn}$ and $\gamma_Q$ are chosen to be of comparable magnitude, so that feasible-region ranking is driven jointly by connectivity and aggregate region quality.

The size score penalises waste:
\begin{equation}
S_\text{size}(R, n) = \begin{cases}
0 & \text{if } |R| < n \\
1 & \text{if } |R| = n \\
e^{-0.5(|R|-n)/n} & \text{if } |R| > n
\end{cases}
\end{equation}
Undersized QVM atomic regions receive a score of 0 (i.e., cannot host the circuit). Exact-fit regions receive the maximum score. Oversized regions are penalised exponentially: a region 50\% larger than needed scores $e^{-0.25} \approx 0.78$; one twice the needed size scores $e^{-0.5} \approx 0.61$. This encourages the use of appropriately sized regions while allowing oversized allocations when necessary.

Allocation proceeds atomically as follows: (1) scan available QVM atomic regions, computing fitness for each; (2) select the region with the highest fitness; (3) mark the selected region as busy; (4) return a mapping from logical qubits (0 to $n-1$) to physical qubits within the region. Upon program completion, the QVM atomic region returns to the available pool. When no single available region is sufficient, \system can also invoke fragmentation-aware allocation extensions and deferred admission policies; these mechanisms are described in Appendix~\ref{app:alloc-extensions}.

\subsection{Hardware-Aware Compilation}
\label{sec:hardware}

After QVM region allocation, \system performs hardware-aware compilation by transpiling each tenant circuit onto its assigned execution region. The core mechanism is \emph{subgraph-restricted compilation}: for a selected region $R$ with vertex set $V_R$ and induced coupler set $E_R$, \system constructs the induced coupling graph $G_R=(V_R,E_R)$ of the QVM region and invokes the Qiskit transpiler with (i) a restricted coupling map corresponding to $G_R$ and (ii) an initial layout that maps logical qubits exclusively onto $V_R$. As a result, all routing, SWAP insertion, and scheduling decisions are confined to the allocated region by construction, providing strict spatial isolation between concurrently executing programs.

\subsubsection{Leveraging and reshaping existing compilation heuristics}
\system deliberately builds on existing compilation stacks, such as Qiskit, rather than introducing a bespoke compiler. Modern quantum transpilers, such as SABRE-style SWAP-based routing~\cite{Li2019,Zou2024}, implement sophisticated heuristic algorithms. However, these fundamentally local heuristics operating over the large combinatorial search space of the latest hundred-qubit scale NISQ processors can produce undesirable behaviour: traversing low-quality/high-crosstalk couplers simply because they offer shorter topological distance, or becoming trapped in locally optimal SWAP patterns.

Subgraph restriction reshapes this optimisation problem rather than replacing it. By limiting the coupling map to $G_R$, \system enforces a \emph{quality-aware prior}. At the same time, the reduced graph can improve both compilation stability and the routing heuristic's solution quality by ensuring that it does not explore noisy or interference-prone areas of the chip. In effect, region-level compilation transforms a difficult global optimisation problem into a smaller, better-conditioned local one, allowing existing heuristics to perform closer to their intended operating regime.

\subsubsection{Virtualisation}
A central architectural property of \system is the explicit decoupling between circuit compilation and resource virtualisation. Virtualisation is realised by materialising each offline-derived QVM region as a self-contained \emph{virtual quantum device}. This region-level hardware view is passed to the transpiler unchanged, where it is treated as the complete target device.

As a result, each circuit is compiled independently against its own virtual device abstraction, with no visibility into other regions, tenants, or global scheduling decisions. Crucially, virtualisation is enforced not by modifying the compiler or embedding tenant-aware logic, but by constraining the compiler’s hardware view to a region-specific coupling map and qubit index space. The transpiler operates on a stable, explicitly defined topology with fixed qubit identifiers, while the execution layer maintains the mapping between QVM region-local qubits and backend-global qubits as metadata.

This design mirrors classical virtualisation in hypervisor-based systems: guest operating systems are not rewritten to enforce isolation; instead, they execute against a restricted and virtualised view of the underlying hardware.

\subsubsection{Region-level compilation}
Although circuits are compiled independently, \system executes them as a single composite backend job, enabling multiprogramming. To bridge this gap, \system maintains explicit mapping metadata that composes region-level compilations into a global execution context. For each circuit $C_i$ assigned to QVM region $R_i$, transpilation produces a circuit expressed over the physical qubits in $V(R_i)$. \system embeds these region-level circuits into a composite circuit by applying a deterministic remapping from local QVM region qubit identifiers to global backend qubit indices. When QVM regions correspond to contiguous index sets, this remapping reduces to a simple offset; in the general case, \system records a permutation map $\pi_i$ that specifies the placement of logical qubits onto physical qubits within $R_i$.

This explicit mapping enables correct and efficient demultiplexing of results. Measurement outcomes returned by the backend are indexed by global physical qubits; \system recovers each tenant’s results by applying the inverse mapping $\pi_i^{-1}$ to extract and reorder the measurement bits corresponding to $R_i$. Because QVM regions are disjoint by construction, composite execution introduces no additional routing constraints, and no re-transpilation is required after composition.

Beyond enforcing spatial separation, region-level compilation reduces opportunities for inter-tenant interference during compilation itself. Scheduling decisions, SWAP insertion, and measurement placement are all made within regions that were selected to minimise crosstalk at their boundaries. This property is particularly important in multi-tenant settings, where simultaneous gate execution and measurement activity can exacerbate control-line and readout crosstalk.

More broadly, the subgraph-restricted compilation framework provides a natural interface for future extensions. Noise-adaptive compilation, region-specific scheduling policies, or dynamic recompilation in response to calibration updates can all be implemented at the region level without entangling compilation logic with global virtualisation or scheduling concerns. In this sense, \system elevates compilation to a first-class, region-scoped service within the quantum virtual machine abstraction.

\section{Evaluation}
\label{sec:eval}
This section evaluates the central empirical claims of \system: improved execution quality, robustness to transient defects, stable multi-tenant batching, and portability across architectures. We first describe the experimental methodology and then report results on offline discovery, simulation across five IBM (heavy-hex) backends, real-device validation on IBM Kingston and Torino, service-level batching behaviour and cost efficiency, cross-architecture validation on Rigetti Ankaa-3, and finally computational overhead. Circuit-level head-to-head and win/loss analyses are deferred to Appendix~\ref{app:circuit}, and detailed timing breakdowns to Appendix~\ref{app:overhead}.

\subsection{Experimental Methodology}
\label{sec:exp-method}

We evaluate performance using 29 circuits from QASMBench~\cite{Li2023}, covering quantum algorithms, arithmetic, variational circuits, error-correction routines, quantum simulation, entanglement generation, and communication-style programs. Circuit widths range from 2 to 10 qubits, with transpiled depths ranging from shallow to moderately deep NISQ workloads. These circuits are evaluated on several backends, including both simulated models and production devices.

For each backend, we first run \system's offline discovery phase to derive QVM atomic regions from the calibration-aware hardware graph. Circuits are then allocated using the best-fit policy and compiled onto the selected regions. We compare against a baseline that uses Qiskit's default backend-aware transpilation flow without \system's quality-aware region discovery and placement. Unless otherwise stated, each circuit is executed with 1024 shots.

We evaluate output quality using the $L1$ distance, Eq.~(\ref{eqn:dldist}), between the ideal and measured output distributions, together with the normalised output-similarity score
\begin{equation}
S = 1 - \frac{1}{2}D_{L1},
\end{equation}
where $S=1$ indicates a perfect match to the ideal output distribution and $S=0$ indicates maximal disagreement\footnote{ $S$ is derived from distributional distance and should not be confused with gate fidelity or state fidelity in the strict quantum-information sense}. We also report relative improvements, lower-percentile behaviour, and circuit-level win/loss statistics where relevant. Additional platform details, calibration scope, workload construction, and reproducibility notes are deferred to Appendix~\ref{app:full-setup}.

\subsection{Discovery Results}

We first evaluate the offline discovery phase of \system that produces the non-overlapping pool of QVM atomic regions. This phase is critical because it determines both the feasibility of dynamic reconfiguration under changing calibrations and the quality of the regions available to the online allocator.

\begin{figure}[htp]
  \centering
  \Description{Connectivity map of the 156-qubit IBM Kingston quantum processor partitioned into multiple coloured subgraphs. Each coloured region represents a disjoint quantum virtual machine allocation region identified from the device coupling topology.}
  \includegraphics[width=0.95\linewidth]{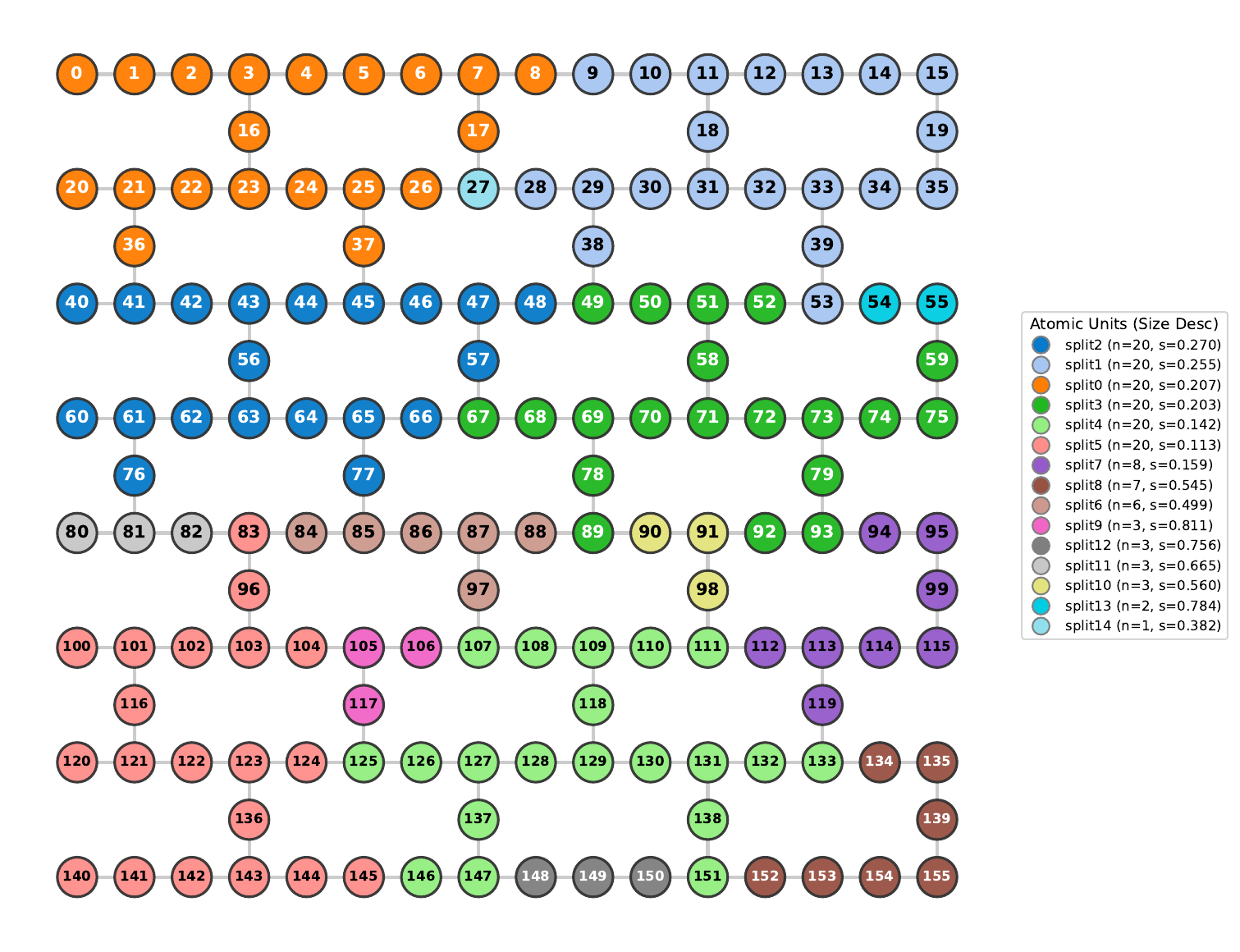}
\caption{\textbf{Visualisation of the offline discovery results on the 156-qubit IBM Kingston backend.} The processor is partitioned into 15 disjoint regions by \system, consisting primarily of QVM atomic regions together with a small number of residual fragments retained to preserve 100\% hardware coverage; the colours indicate their corresponding scores. Unlike rigid geometric partitioning, these irregular subgraphs naturally adapt to the heavy-hex topology and connectivity constraints, ensuring 100\% hardware coverage as quantified in Table~\ref{tab:discovery}.}
  \label{fig:kingston_map}
\end{figure}

\system consistently produces a compact set of non-overlapping execution regions with full hardware coverage and clear quality stratification across all backends. We show a representative  outcome for IBM Kingston (156 qubits, heavy-hex) in  Table~\ref{tab:discovery} and Figure~\ref{fig:kingston_map}.

\begin{table}[t]
\centering
\caption{Region discovery on IBM Kingston (156 qubits)}
\label{tab:discovery}
\small
\begin{tabular}{lc}
\toprule
\textbf{Metric} & \textbf{Value} \\
\midrule
\textbf{Total discovery time} & \textbf{0.81 s} \\
\midrule
Regions discovered & 15 \\
Hardware coverage & 100\% (156/156 qubits) \\
Region size range & 1--20 qubits \\
Mean region size & 10.4 qubits \\
Quality score range & 0.11--0.81 \\
Quality spread (max/min) & 7.2$\times$ \\
\bottomrule
\end{tabular}
\end{table}

The quality-weighted Louvain algorithm partitions the entire coupling graph into a compact set of non-overlapping regions, assigning 153 qubits to 13 execution regions and leaving two residual fragments to preserve full coverage of all 156 qubits. In contrast to geometric or template-based partitioning approaches, which can leave large numbers of single-qubit fragments on heavy-hex topologies because fixed region shapes do not align cleanly with irregular connectivity, \system substantially reduces the probability of stranding qubits at region boundaries. Full coverage ensures that all available hardware resources remain usable and maximises the achievable degree of concurrency in multi-tenant execution.

Critically, these atomic regions retain internal homogeneity, while dealing with the spatial heterogeneity of the backend. For a backend like IBM Kingston with significant heterogeneity, these regions exhibit pronounced variation in execution quality. Here, region scores span a 7.2$\times$ range between the highest- and lowest-scoring regions, reflecting region-level differences arising from the joint effects of gate fidelity, readout quality, connectivity, and internal uniformity.

This stratification has direct operational implications. High-quality regions consistently feature low-error couplers and support reliable execution of noise-sensitive workloads, whereas lower-scoring regions correspond to noisier areas of the processor and are better suited for noise-insensitive, exploratory, or test circuits. 

The discovery phase completes in 0.81 seconds on a 156-qubit device on a GH200-based platform, fast enough to be re-executed at each calibration update. As subgraph construction, community detection, and scoring scale linearly with graph size, this approach remains practical even for thousands of qubits. A detailed timing breakdown and an additional discussion of overhead are deferred to Appendix~\ref{app:overhead}.

\subsection{Simulation Results on Five IBM Backends}
\label{sec:multibackend}
We evaluate five simulated IBM heavy-hex backends: Pittsburgh, Kingston, Fez, Torino, and Marrakesh, to assess performance robustness under diverse noise characteristics and degrees of hardware heterogeneity. We use the default compiler to compile all 29 circuits.

\begin{figure}[t]
  \centering
  \begin{subfigure}[b]{0.48\textwidth}
    \centering
    \Description{Boxplot chart showing distributions of total variation ($L1$) error for benchmark quantum circuits across several simulated hardware backends. We compare the default compiler to \system.}
    \includegraphics[width=\linewidth, keepaspectratio]{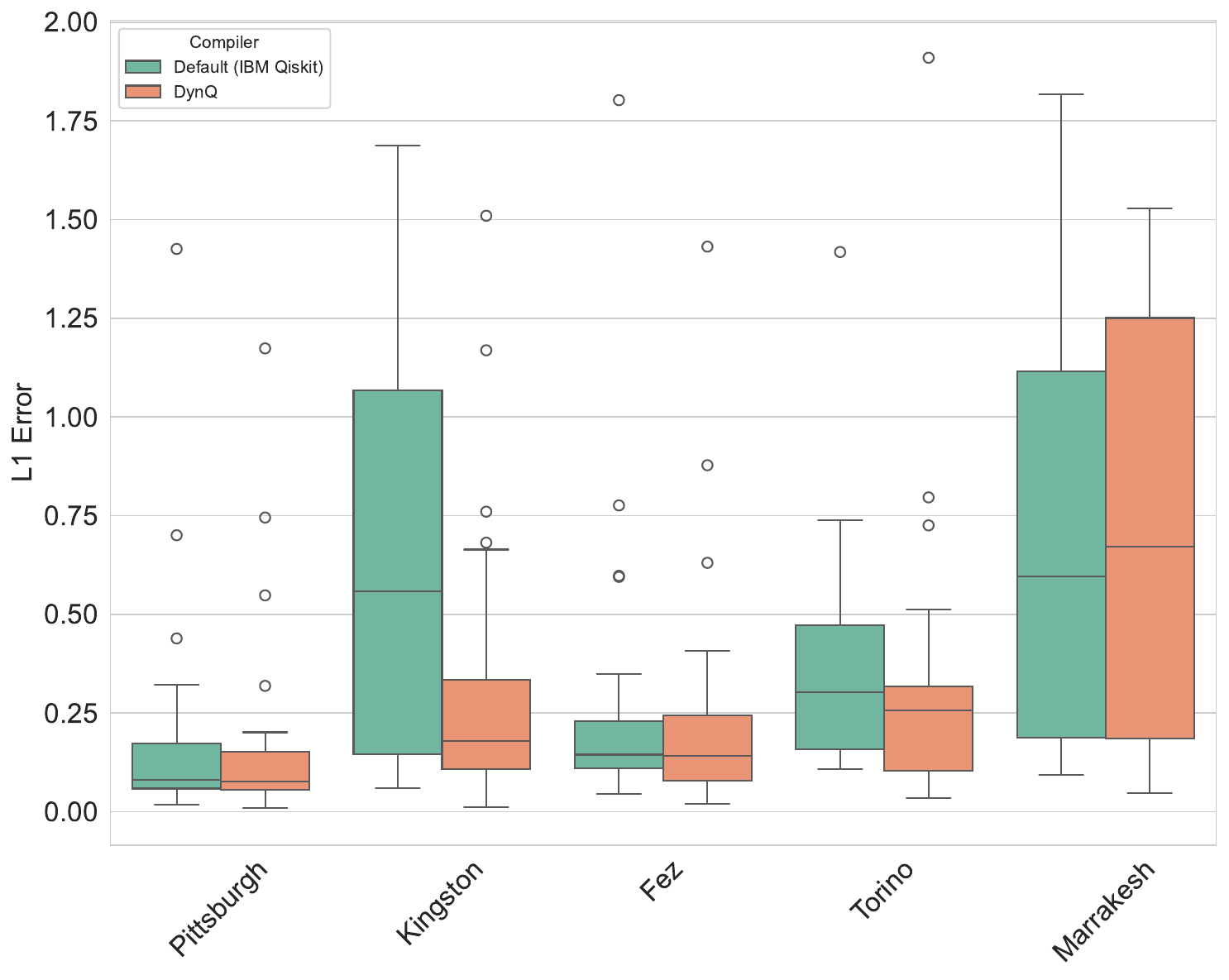}
    \caption{$L1$ Error Distribution (Simulated Backends)}
    \label{fig:l1_error_dist}
  \end{subfigure}
  \hfill 
  \begin{subfigure}[b]{0.48\textwidth}
    \centering
    \Description{Boxplot chart showing distributions of circuit output similarity for benchmark quantum circuits across several simulated hardware backends. }
    \includegraphics[width=\linewidth, keepaspectratio]{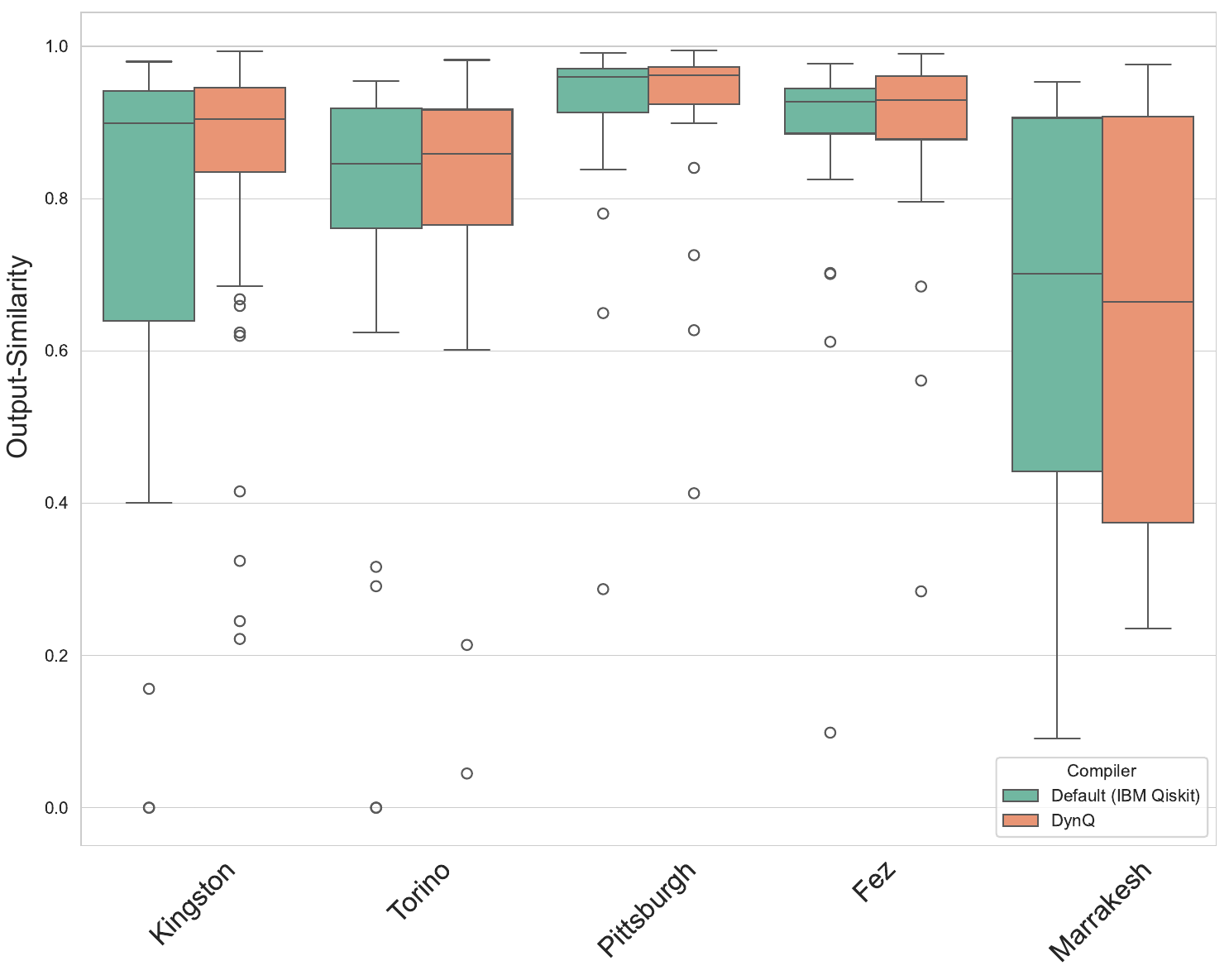}
    \caption{Output-similarity Distribution  (Simulated Backends)}
    \label{fig:fidelity_dist}
  \end{subfigure}

  \caption{\textbf{Performance characterisation of the \system compiler under noisy simulation across diverse backend landscapes.}
  (a) \textbf{$L1$ Error Distribution:} The Total Variation Distance ($L1$ Error) between the ideal and noisy output probability distributions. Lower values indicate higher accuracy. \system (orange) consistently exhibits lower error rates compared to the default compiler (green) across all simulated chips.
  (b) \textbf{Output-similarity Distribution:} Circuit output-similarity scores for the same set of benchmarks. \system demonstrates higher median output-similarity and lower performance variance, thereby mitigating noise-induced errors.}
  \label{fig:performance_distribution}
\end{figure}

Figure~\ref{fig:performance_distribution} visualises the resulting $L1$ error and output-similarity distributions across backends, highlighting how \system consistently shifts both metrics toward lower error and higher output-similarity regimes while reducing performance variance as summarised in  Table~\ref{tab:multibackend}.

Across all platforms, \system consistently improves the average output similarity $S$ and reduces the $L1$ distance relative to the standard compilation baseline. The magnitude of improvement, however, varies significantly across devices, reflecting differences in the underlying noise landscape. Importantly, the benefit of \system is not captured fully by raw win rate alone: the gains are asymmetric, with large improvements concentrated in challenging cases and only moderate regressions elsewhere; detailed head-to-head statistics are reported in Appendix~\ref{app:circuit}.

\begin{table}[!htp]
\centering
\caption{Aggregate output-similarity and $L1$ error comparison across five simulated IBM Quantum backends. \system achieves the largest relative gains on backends with high noise variance (notably Kingston), significantly raising the performance floor (as indicated by P25 output-similarity).}
\label{tab:multibackend}
\small
\setlength{\tabcolsep}{4pt}
\begin{tabular}{llcccccc}
\toprule
\multirow{2}{*}{\textbf{Backend}} 
& \multirow{2}{*}{\textbf{Method}} 
& \multicolumn{3}{c}{\textbf{Similarity}} 
& \multicolumn{2}{c}{\textbf{$L1$ Error}} 
& \textbf{Rel.\ $L1$} \\
\cmidrule(lr){3-5} \cmidrule(lr){6-7}
& & \textbf{Mean} & \textbf{Median} & \textbf{P25} & \textbf{Mean} & \textbf{Median} & \textbf{Red.} \\
\midrule
\multirow{2}{*}{Pittsburgh}
& Baseline & 0.910 & 0.960 & 0.913 & 0.180 & 0.080 & -- \\
& \system  & \textbf{0.917} & \textbf{0.962} & \textbf{0.924} & \textbf{0.166} & \textbf{0.076} & 7.8\% \\
\midrule
\multirow{2}{*}{Fez}
& Baseline & 0.869 & 0.928 & \textbf{0.886} & 0.262 & 0.145 & -- \\
& \system  & \textbf{0.884} & \textbf{0.930} & 0.878 & \textbf{0.232} & \textbf{0.141} & 11.5\% \\
\midrule
\multirow{2}{*}{Torino}
& Baseline & 0.825 & 0.849 & 0.764 & 0.351 & 0.303 & -- \\
& \system  & \textbf{0.844} & \textbf{0.872} & \textbf{0.841} & \textbf{0.311} & \textbf{0.256} & 11.2\% \\
\midrule
\multirow{2}{*}{Kingston}
& Baseline & 0.702 & 0.721 & 0.466 & 0.595 & 0.558 & -- \\
& \system  & \textbf{0.837} & \textbf{0.911} & \textbf{0.833} & \textbf{0.327} & \textbf{0.179} & \textbf{45.1\%} \\
\midrule
\multirow{2}{*}{Marrakesh}
& Baseline & 0.638 & \textbf{0.702} & \textbf{0.442} & 0.723 & \textbf{0.597} & -- \\
& \system  & \textbf{0.647} & 0.664 & 0.375 & \textbf{0.707} & 0.672 & 2.3\% \\
\bottomrule
\end{tabular}
\end{table}

\paragraph{Impact of Hardware Heterogeneity}

The effectiveness of \system is strongly correlated with the degree of spatial noise heterogeneity on the target backend. On \textbf{Kingston}, which exhibits pronounced variance in two-qubit gate errors across the chip, \system achieves a substantial 19.1\% relative improvement in the output-similarity score and a 45.1\% reduction in $L1$ error. This behaviour reflects the ability of quality-aware region selection to systematically avoid low-quality subgraphs and route circuits toward favourable hardware regions.

In contrast, \textbf{Pittsburgh} is characterised by uniformly high-quality qubits, with baseline output-similarity already exceeding 0.91. In this regime, optimisation headroom is inherently limited, and \system yields small but consistent gains (+0.8\%) without introducing regressions. Similarly, \textbf{Marrakesh}, despite high absolute noise, exhibits relatively uniform error characteristics across the device. As a result, selective placement offers limited leverage, and the observed improvement remains modest (+1.3\%).

These results indicate that \system is most effective in environments with \emph{exploitable heterogeneity}, where meaningful quality gradients exist across the hardware graph, rather than in uniformly good or uniformly poor regimes.

At the circuit level, these gains are not uniformly distributed. They are concentrated primarily in low-output-similarity baseline regimes, where quality-aware placement can avoid severely degraded subgraphs and recover otherwise poor executions, while remaining close to performance-neutral on already strong mappings. A detailed distributional and circuit-level analysis is deferred to Appendix~\ref{app:circuit}.

\subsection{Real-Device Validation on IBM Kingston and Torino}
\label{sec:realdevice}

We next evaluate \system on production IBM Quantum hardware to assess whether the gains observed under calibration-derived noise simulation translate to real executions that include additional device effects, such as drift, coherent miscalibration, and time-varying operational constraints. We execute the same 29-circuit QASMBench workload on two heavy-hex backends, IBM Kingston (156 qubits) and IBM Torino (133 qubits).

\begin{figure}[t]
  \centering
  \begin{subfigure}[b]{0.48\textwidth}
    \centering
    \Description{Boxplot chart showing distributions of total variation ($L1$) error for benchmark quantum circuits executed on real IBM Quantum processors. Results compare \system and the default compiler across multiple devices.}
    \includegraphics[width=\linewidth, keepaspectratio]{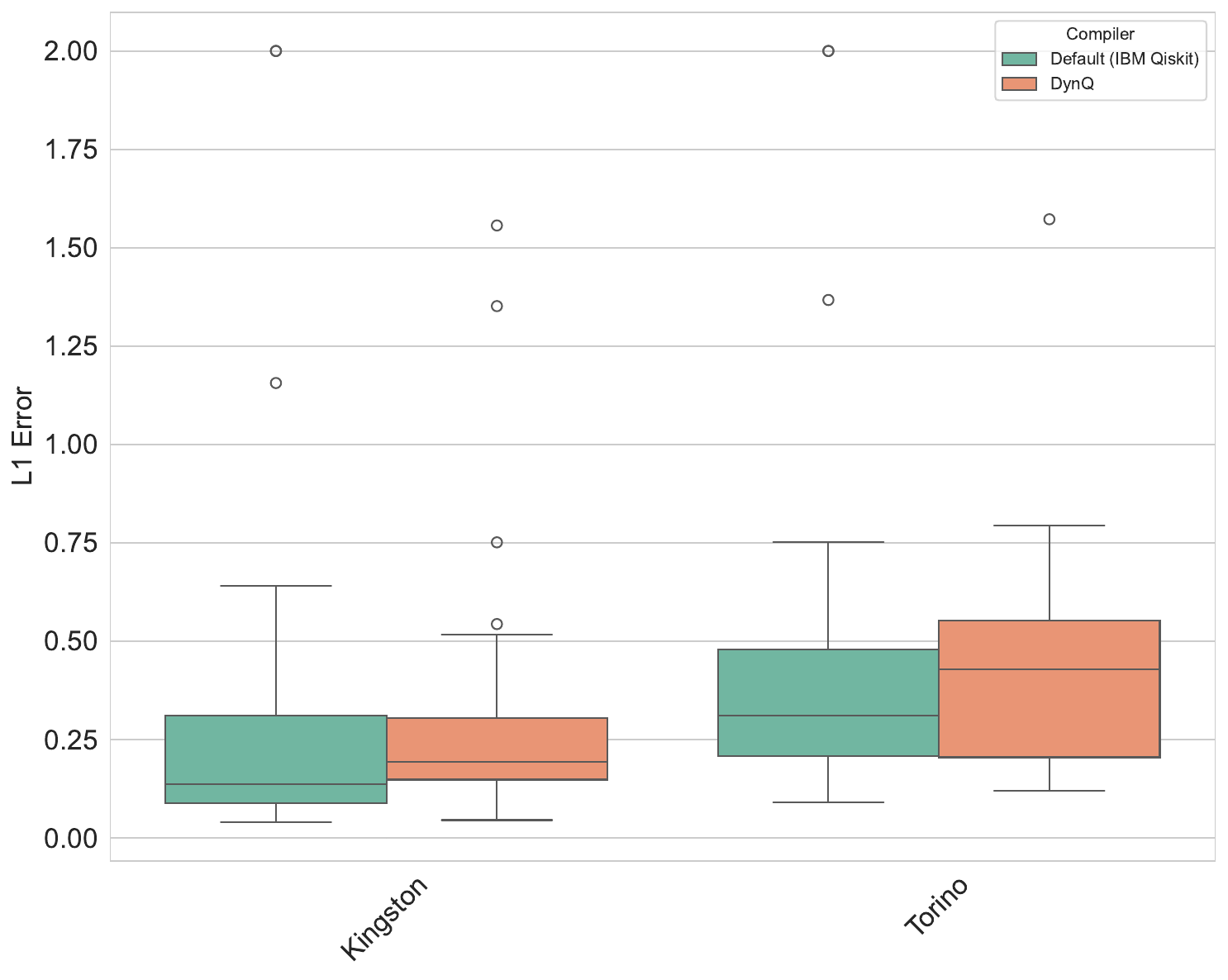}
    \caption{$L1$ Error Distribution (Real Device)}
    \label{fig:real_l1_error}
  \end{subfigure}
  \hfill 
  \begin{subfigure}[b]{0.48\textwidth}
    \centering
    \Description{Boxplot chart showing distributions of circuit output similarity for benchmark quantum circuits executed on real IBM Quantum processors.}
    \includegraphics[width=\linewidth, keepaspectratio]{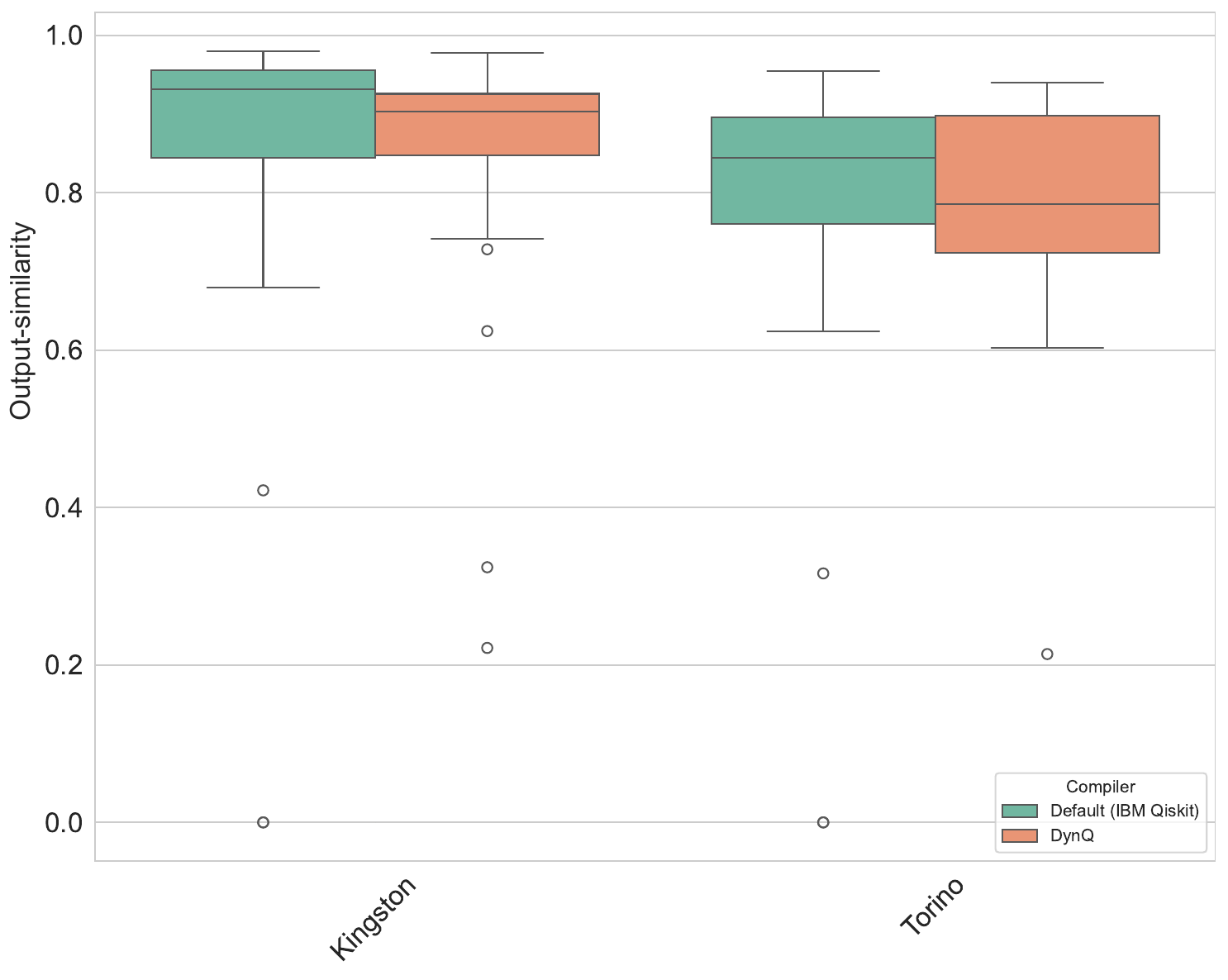}
    \caption{Output-Similarity Distribution (Real Device)}
    \label{fig:real_fidelity}
  \end{subfigure}

    \caption{\textbf{Experimental validation of \system on real IBM Quantum processors.}
    (a) \textbf{$L1$ error distribution:} the total variation distance between the measured and ideal output distributions for the 29-circuit workload executed on IBM Kingston and Torino. Lower values indicate better agreement with the ideal distribution.
    (b) \textbf{Output-similarity distribution:} the corresponding normalised similarity score $S$. Relative to the default compiler, \system shifts the distribution toward higher scores, tightens the lower tail, and reduces the incidence of near-failure executions.}
  \label{fig:real_device_performance}
\end{figure}

Figure~\ref{fig:real_device_performance} presents the resulting distributions of $L1$ error and normalised output-similarity score across both devices, capturing execution-level robustness beyond average-case metrics. Consistent with the simulation results, \system shifts the distributions toward lower-error, higher-score regions while substantially reducing performance variance and mitigating near-failure executions. Table~\ref{tab:device} reports the corresponding aggregate output quality using the normalised similarity score $S = 1-\tfrac{1}{2}D_{L1}$ and the associated $L1$ distance $D_{L1}$. Detailed circuit-level, regression, and head-to-head analyses are deferred to Appendix~\ref{app:circuit}.

\begin{table}[t]
\centering
\caption{Real device results on IBM Kingston (156 qubits) and IBM Torino (133 qubits) for the 29-circuit QASMBench workload.}
\label{tab:device}
\small
\begin{tabular}{llcccc}
\toprule
\textbf{Backend} & \textbf{Method} & \textbf{Mean $S$} & \textbf{Std($S$)} & \textbf{Mean $D_{L1}$} & \textbf{Failures} \\
\midrule
\multirow{2}{*}{Kingston} 
& Baseline & 0.823 & 0.258 & 0.355 & 2 \\
& \system  & \textbf{0.840} & \textbf{0.177} & \textbf{0.319} & \textbf{0} \\
\midrule
\multirow{2}{*}{Torino} 
& Baseline & 0.762 & 0.245 & 0.475 & 2 \\
& \system  & \textbf{0.787} & \textbf{0.150} & \textbf{0.426} & \textbf{0} \\
\bottomrule
\end{tabular}
\end{table}

\paragraph{Aggregate gains and variance reduction.}
Across both devices, \system improves mean output quality while also reducing cross-circuit variability across the workload. On Kingston, mean output-similarity increases from $0.823$ to $0.840$ (a $2.2\%$ relative gain), while mean $L1$ error decreases from $0.355$ to $0.319$ (a $10.1\%$ reduction). On Torino, which has a lower baseline output similarity, \system yields a larger relative uplift, improving mean output-similarity by $3.2\%$ and reducing mean $L1$ error by $10.2\%$. Notably, the standard deviation of output-similarity drops substantially (Kingston: $32\%$, Torino: $39\%$), indicating that quality-aware QVM region selection improves not only average performance but also the \emph{predictability} of outcomes across heterogeneous circuits.

\paragraph{Per-circuit heterogeneity and ``floor-raising'' behaviour.}
Per-circuit outcomes remain diverse on real hardware. On Kingston, $11/29$ circuits improve under \system and $18/29$ decrease slightly; on Torino, the split is $12/29$ versus $17/29$. Despite a win rate below $50\%$, the overall mean improves, implying that \system's \emph{large positive recoveries} outweigh its typically smaller regressions. This behaviour is consistent with \system acting as a floor-raising mechanism: it is most beneficial when the baseline mapping lands on locally poor hardware. In particular, \system recovers two baseline ``failures'' on each backend, where the baseline execution collapses to $S\approx 0$ (equivalently $D_{L1}=2$). On Kingston, \texttt{bell\_n4} and \texttt{qaoa\_n3} improve from $S=0.00$ to $S=0.906$ and $S=0.860$, respectively; on Torino, the same circuits are recovered from $S=0.00$ to $S=0.879$ and $S=0.647$. These cases materially improve user-visible service quality by converting paid-but-unusable runs into successful executions.

\paragraph{Structured regressions and baseline-dependent gains.}
Regressions are concentrated in circuits that already achieve near-ceiling output similarity under baseline placement, where dynamic remapping can sacrifice an already favourable configuration by introducing additional routing overhead or exposing the circuit to a less benign error structure. More broadly, the gains on real hardware remain strongly dependent on the baseline mapping, with the largest improvements occurring precisely when the default mapping performs poorly. Detailed circuit-level, head-to-head, and regression analyses are deferred to Appendix~\ref{app:circuit}.

\paragraph{Simulation versus device.}
Finally, the absolute improvement magnitudes differ between the simulation and the device for some circuits, reflecting error mechanisms that standard calibration-derived noise models do not fully capture. Simulators predominantly capture incoherent processes (e.g., relaxation and stochastic gate errors), whereas real devices also exhibit coherent effects (systematic over-rotations, drift, and residual interactions) that can alter a given circuit's sensitivity to remapping and routing choices. The fact that \system provides consistent aggregate gains on both Kingston and Torino indicates that the approach remains effective under this broader, more realistic error landscape. 

\subsection{Service-Level Scalability and Cost Efficiency}
\label{sec:service}
\begin{figure}[!htp]
  \centering
  \Description{Line plot showing average circuit output-similarity on the IBM Kingston quantum processor as a function of batch size during concurrent execution. The curve remains approximately stable as job density increases, with a shaded band indicating variability.}
  \includegraphics[width=0.85\linewidth]{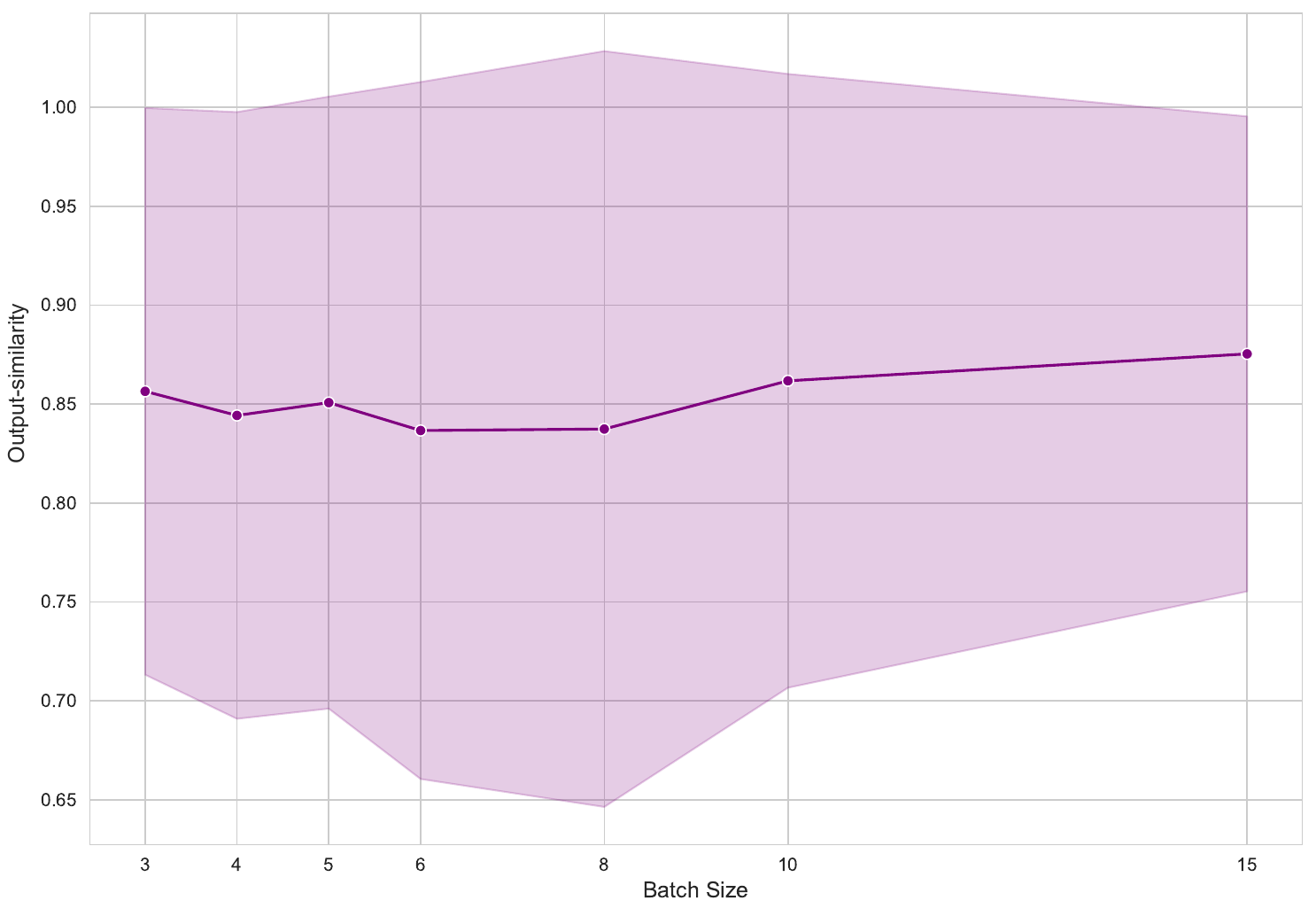}
  \caption{\textbf{Robustness against crosstalk and congestion in batched execution.} 
  The average output similarity as a function of batch size on the Kingston device, with the standard deviation shown in the shaded region. 
  This stability evidences \system's effectiveness in crosstalk-aware routing and spatial isolation during multi-program compilation. 
  }
  
  \label{fig:batch_stability}
\end{figure}

We evaluate how \system performs as workload concurrency increases using the Kingston backend with batch sizes ranging from 2 to 18 programs. The results are summarised in Table~\ref{tab:batch} and Figure~\ref{fig:batch_stability}.

\begin{table}[t]
\centering
\caption{Batch scalability on IBM Kingston}
\label{tab:batch}
\small
\begin{tabular}{cccc}
\toprule
\textbf{Batch Size} & \textbf{Regions Used} & \textbf{Mean Output Similarity} & \textbf{Std Dev} \\
\midrule
2 & 2 & 87.5\% & 0.120 \\
3 & 3 & 86.2\% & 0.155 \\
4 & 4 & 83.7\% & 0.191 \\
5 & 5 & 83.7\% & 0.176 \\
6 & 5 & 86.5\% & 0.138 \\
7 & 5 & 83.6\% & 0.171 \\
8 & 4 & 84.4\% & 0.153 \\
10 & 3 & 87.1\% & 0.130 \\
15 & 3 & 84.5\% & 0.146 \\
18 & 3 & 85.3\% & 0.156 \\
\bottomrule
\end{tabular}
\end{table}

This table and figure show that output-similarity {\em{remains stable}} as batch size increases. The correlation coefficient between batch size and output-similarity is $r = -0.012$, indicating an effectively zero linear relationship within measurement noise. This stability is remarkable: on superconducting quantum chips, increasing qubit activation density typically causes crosstalk degradation from frequency collisions and $ZZ$ coupling between neighbouring qubits. Accordingly, output-similarity would generally be expected to decrease with batch size.

This behaviour is consistent with \textbf{implicit crosstalk suppression} induced by the partitioning and allocation process.  Region boundaries are placed along high-error edges, which often correspond to physically distant or frequency-detuned qubit pairs where crosstalk is naturally lower. Additionally, the allocation algorithm distributes programs across non-adjacent regions when feasible, thereby creating buffer zones that further reduce inter-program interference. 

This batch stability, along with reduced execution failures, has significant implications for the economics of quantum cloud computing. \system allows more programs to be packed onto a processor without sacrificing per-program output-similarity, yielding a linear increase in utilisation at constant quality, while also improving effective availability.

We analyse the cost and throughput implications in 
Table~\ref{tab:cost}. Sequential execution requires 29 independent jobs. Under \system, batching two circuits per job reduces the required number of jobs to 15, corresponding to a 48\% reduction in job count and a 1.9$\times$ throughput gain, while maintaining a mean output similarity of 87.5\%. As the batch size increases, the economic benefits scale rapidly. With 10 circuits per batch, the entire workload completes in just 3 jobs, achieving a 90\% reduction in job count and a 9.7$\times$ increase in throughput, while preserving a mean output-similarity of 87.1\%. From a service perspective, the combined effect of failure elimination and safe batching is a substantial reduction in the cost per successful computation.

\begin{table}[t]
\centering
\caption{Cost reduction through batching (29-circuit workload)}
\label{tab:cost}
\small
\begin{tabular}{cccccc}
\toprule
\textbf{Batch} & \textbf{Jobs} & \textbf{Cost} & \textbf{Throughput} & \textbf{Mean} \\
\textbf{Size} & \textbf{Needed} & \textbf{Reduction} & \textbf{Gain} & \textbf{Output Similarity} \\
\midrule
1 (baseline) & 29 & -- & 1.0$\times$ & -- \\
2 & 15 & 48\% & 1.9$\times$ & 87.5\% \\
4 & 8 & 72\% & 3.6$\times$ & 83.7\% \\
6 & 5 & 83\% & 5.8$\times$ & 86.5\% \\
10 & 3 & 90\% & 9.7$\times$ & 87.1\% \\
15 & 2 & 93\% & 14.5$\times$ & 84.5\% \\
\bottomrule
\end{tabular}
\end{table}

\subsection{Cross-Architecture Validation on Rigetti Ankaa-3}
\label{subsec:rigetti_evaluation}
\begin{figure}[!htbp]
  \centering
  \Description{Connectivity map of the Rigetti Ankaa-3 quantum processor showing qubits grouped into several coloured regions. Each coloured subgraph represents a region assignment identified from the device coupling topology.}
  \includegraphics[width=0.7\linewidth]{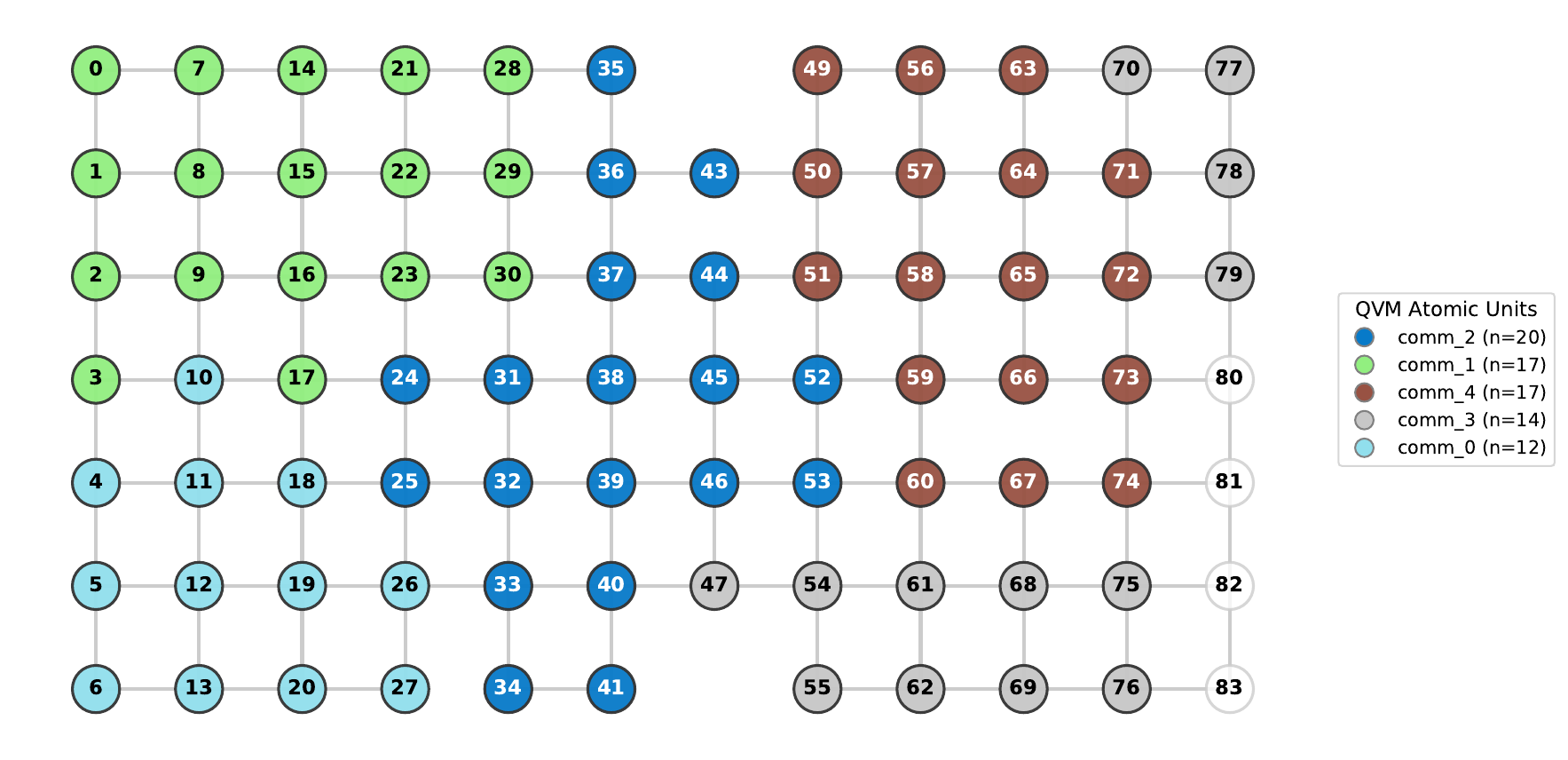}
  \caption{\textbf{Offline discovery results on the Ankaa-3 backend.} The coloured regions show the region assignments identified by \system\ under the provider-compatible execution view. Unlike rigid geometric partitioning, these irregular subgraphs adapt to the underlying hardware topology and connectivity constraints.}
  \label{fig:ankaa3_map}
\end{figure}

\begin{figure}[t]
  \centering
  \Description{Boxplot showing distribution of output-similarity for benchmark circuits executed on Rigetti Ankaa-3. Two groups are shown: the default compiler and \system. \system results exhibit higher minimum output-similarity and narrower interquartile spread.}
  \includegraphics[width=0.78\linewidth]{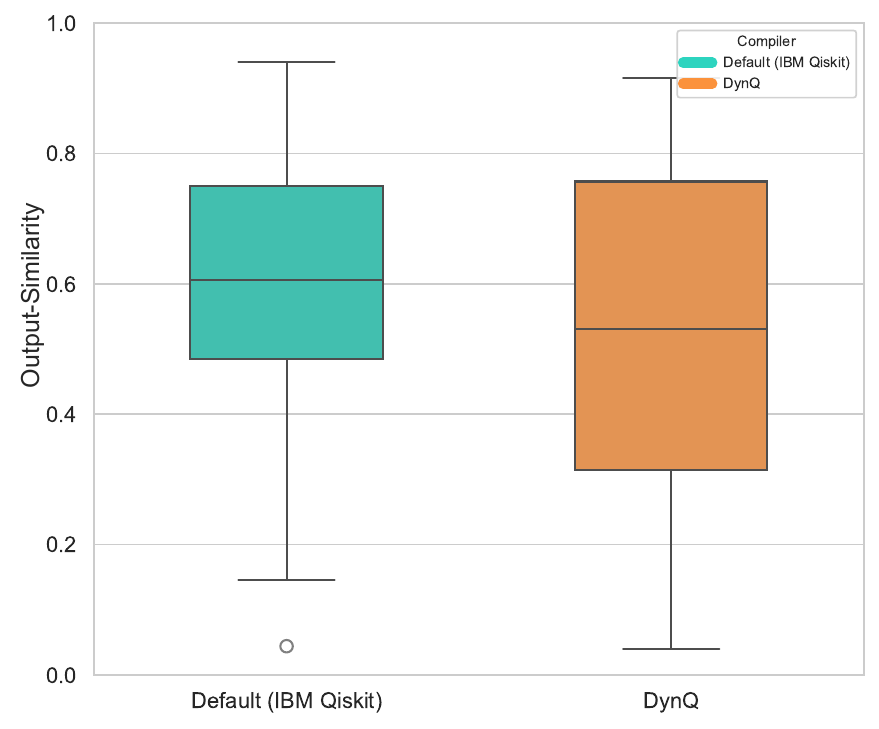}
    \caption{\textbf{Output-similarity Distribution:} Experimental output-similarity scores comparing the strictly sequential default Qiskit compiler (green) against \system's spatial multiplexing approach (orange) on Rigetti Ankaa-3. Unlike the IBM Quantum results, the Ankaa-3 distribution does not show a uniform shift in favour of \system. Instead, the wider spread under \system reflects a circuit-dependent trade-off: some workloads improve when better subgraph allocation reduces routing overhead, while others regress under the backend's dense-lattice, noise-limited operating conditions (see details in Figure~\ref{fig:ankaa-gain}). These results highlight the bounded and hardware-dependent nature of software-level mitigation on early-stage NISQ devices. Note that we use qiskit-braket-provider~\cite{QiskitBraketProvider2025} to compile and submit jobs to AWS Braket.}
  \label{fig:ankaa-boxplot}
\end{figure}

To probe the portability of \system's abstraction layer across different hardware architectures, we extend our evaluation to the Rigetti Ankaa-3 processor~\cite{AWS2025a}, as shown in Figure~\ref{fig:ankaa3_map}. The figure reflects the provider-compatible execution view used in our Braket experiments: unusable qubits 42 and 48 were excluded from \system's effective topology, while a conservative 80-qubit metadata override was required to keep the Qiskit--Braket submission path operational.\footnote{The qiskit-braket-provider path could not faithfully represent partial qubit unavailability under the original 84-qubit backend metadata. We therefore used a conservative 80-qubit provider-visible metadata override to preserve compilability and executability. Any apparent fragmentation in Figure~\ref{fig:ankaa3_map} is a toolchain artefact of this workaround rather than a disconnected QVM atomic region produced by \system.} Unlike the sparse, low-degree heavy-hex lattice characteristic of IBM processors, Ankaa-3 implements an 84-qubit square lattice with tunable couplers. This denser degree-4 connectivity changes the compilation landscape: it can reduce the SWAP overhead required for routing, but it may also increase the exposure of concurrent workloads to hardware-level interference effects. By deploying 29 diverse benchmark circuits via Amazon Braket, we investigate how \system behaves under these competing architectural conditions relative to a strictly sequential, temporally isolated baseline, as summarised in Figure~\ref{fig:ankaa-boxplot}.

\begin{figure}[t]
  \centering
  \Description{Bar chart showing mean output-similarity difference between \system and the default compiler for 29 benchmark quantum circuits executed on the Rigetti Ankaa-3 processor. Bars above zero indicate higher output-similarity achieved by \system, while bars below zero indicate cases where the default compiler performs better. Circuits vary in routing complexity and qubit locality.}
  \includegraphics[width=0.78\linewidth]{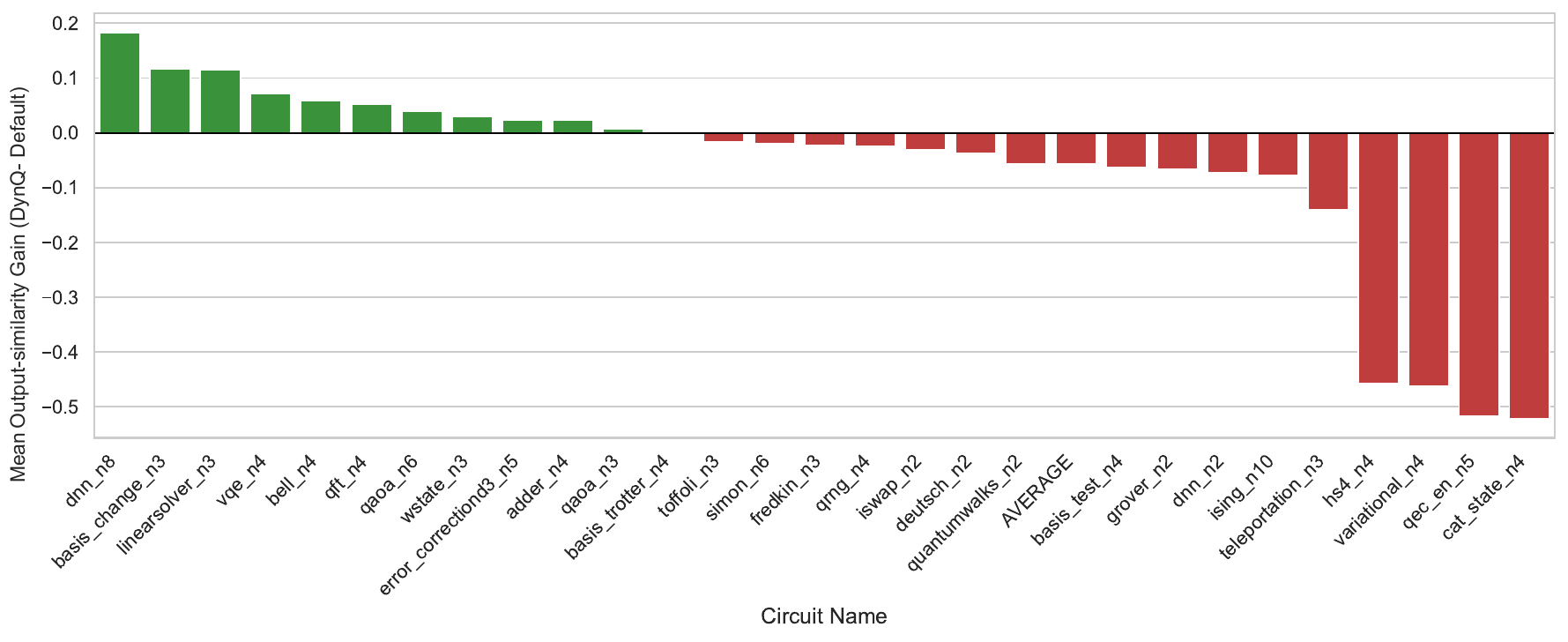}
  \caption{\textbf{Absolute Output-similarity Gain per Circuit:} The difference in mean output-similarity (\system minus Default) across the 29 benchmark circuits. Green bars indicate topology-sensitive circuits (e.g., \texttt{dnn\_n8}, \texttt{basis\_change\_n3}) where the output-similarity preserved by \system's routing optimisation vastly outweighs any introduced multiprogramming noise. Conversely, red bars denote shallow, localised circuits (e.g., \texttt{cat\_state\_n4}) where routing benefits are marginal, rendering them slightly more susceptible to the residual hardware crosstalk inherent in concurrent execution on a dense square lattice.}
  \label{fig:ankaa-gain}
\end{figure}

\paragraph{Dense-lattice trade-offs.}
Evaluating QVM allocation on near-term hardware requires acknowledging the absolute physical limitations of the underlying backend. On Ankaa-3, the baseline output similarity of several circuits is already severely constrained by the intrinsic noise floor, with some workloads falling well below 0.5 (e.g., \texttt{basis\_change\_n3} at 0.150 and \texttt{dnn\_n8} at 0.275). In such coherence-limited regimes, circuit execution is often dominated by global decoherence ($T_1$/$T_2$ relaxation), native two-qubit gate infidelities, and a more complex hardware noise model, rather than by topology-induced routing overhead alone. While \system can optimise logical-to-physical placement to exploit higher-quality processor regions, software-level remapping cannot fundamentally recover quantum information that has already dissipated into the environment. Consequently, the achievable absolute gain is physically bounded by the backend's ambient error rates, illustrating the limits of software mitigation on early-stage NISQ hardware.

\paragraph{Utilisation implications.}
Beyond output similarity, the clearest architectural contribution of \system on Ankaa-3 lies in mitigating spatial fragmentation. Under a sequential execution model, small-to-medium circuits (e.g., 2- to 10-qubit workloads) monopolise the full 84-qubit QPU while leaving most coherent resources idle. \system instead converts this idle capacity into concurrent throughput by partitioning the lattice into calibration-derived regions and compacting the benchmark set into a much smaller set of execution batches. Even when fidelity gains are mixed, this multi-tenancy mechanism remains operationally valuable in cloud settings: it amortises submission, queueing, and execution overheads across multiple workloads, reducing the effective cost of serving small jobs on a large device. Taken together, these results suggest that topology-agnostic virtualisation is valuable on dense-lattice hardware primarily as a utilisation and cost-amortisation mechanism, while fidelity gains remain workload-dependent and bounded by backend quality.

\subsection{Computational Overhead}
The computational cost of \system is dominated by the offline discovery phase, which runs once per calibration cycle and completes in under 1 second. Online allocation remains negligible at 0.08 ms per circuit, which is four orders of magnitude faster than circuit execution. The primary runtime overhead beyond allocation is transpilation on cache miss, which is dominated by Qiskit's backend optimiser and would occur regardless of the partitioning approach. Cache hits reduce this to 0.1 ms for repeated circuit structures. Detailed timing breakdowns and additional sensitivity discussion are reported in Appendix~\ref{app:overhead}.

\section{Discussion}

\subsection{Why Community Detection is Effective for Quantum Partitioning}

Our partitioning strategy rests on a critical but straightforward observation: errors on contemporary quantum processors are not spatially independent. Instead, calibration-derived quality metrics, such as two-qubit gate error rates and readout errors, exhibit strong spatial locality. Qubits and couplers that are physically or electrically proximal tend to share similar noise characteristics, forming contiguous regions of relatively uniform quality across the device.

This spatial structure arises from several overlapping mechanisms. 
\emph{Process-induced variations} during fabrication introduce wafer-scale gradients and local non-uniformities, leading to smooth spatial variation in qubit frequencies, anharmonicities, and calibration stability. 
\emph{Spectral crowding and frequency management}, particularly on sparse topologies such as heavy-hex, further accentuate this effect: regions with dense spectral packing are more prone to frequency collisions and parasitic interactions, producing sharp quality discontinuities aligned with the coupling graph. 
In addition, \emph{shared environmental perturbations}, including substrate defects, magnetic fluctuations, and thermal gradients, can induce correlated noise over neighbourhoods of qubits, creating extended areas of consistently higher or lower output-similarity.

Taken together, these effects imply that a quantum processor can be naturally modelled as a \emph{quality-weighted interaction graph} with meaningful community structure. High-fidelity couplers form densely connected subgraphs, while unreliable or interference-prone couplers appear as weak links between them. Community detection is therefore well matched to the partitioning problem: by optimising modularity on a graph weighted by current calibration data, the algorithm aggregates qubits connected by consistently reliable interactions and places region boundaries along persistently weak couplers. Crucially, this process does not require explicit modelling or attribution of individual error sources; it exploits their combined effect as directly reflected in calibration measurements.

The detailed circuit-level analysis further shows that the largest gains arise for narrow circuits and for executions whose baseline placement traverses severely degraded hardware regions, whereas regressions are concentrated in width-limited regimes where the best discovered regions are too small for the requested circuit (Appendix~\ref{app:circuit}).

\subsection{Robustness to Transient Link Failures}

Real-device runs also highlight an operational robustness property of \system, which we refer to as \emph{dead-link immunity}. In practice, couplers may become temporarily unavailable or effectively unusable due to calibration failures, transient defects, or elevated error rates. Compilers that assume a largely static coupling map can inadvertently route through such links, leading to pathological transpilation behaviour or outright execution failure.

\system mitigates this failure mode by constructing the routing graph from up-to-date calibration information. Couplers that are flagged as failed are removed; couplers whose quality degrades beyond thresholds are assigned negligible weights. QVM region discovery, therefore, tends to exclude these links from region interiors, and subsequent routing and allocation steps avoid relying on them as critical pathways. As a result, \system more consistently targets \emph{currently viable} connectivity, improving operational availability under time-varying hardware conditions. This robustness is particularly relevant in shared, cloud-style settings, where transient hardware anomalies can otherwise translate into user-visible failures.

\subsection{Crosstalk Suppression}

In our batch scalability experiments, output similarity remained stable as the number of concurrently executed programs increased from 2 to 18. This behaviour is non-trivial: higher concurrent activity typically increases crosstalk and spectator effects, thereby degrading performance.

We hypothesise that the observed stability is primarily driven by \emph{implicit spatial isolation} induced by the partitioning process. Because region boundaries preferentially align with low-weight couplers, the cut edges often correspond to qubit pairs that are (i) physically more separated (e.g., longer effective coupling paths), (ii) more frequency-detuned (reducing residual $ZZ$-type interactions), or (iii) empirically prone to crosstalk as reflected in elevated calibrated error rates. Concretely, low-weight edges are natural candidates for workload separation because they are precisely where inter-region interference is likely to be most damaging.

By placing QVM partition boundaries at these weak links, \system effectively introduces buffer zones between concurrently active QVM regions. The allocation policy further amplifies this effect by preferentially distributing programs across non-adjacent regions when capacity permits. Together, these mechanisms suggest that quality-aware community partitions can serve as a practical substrate for multiprogrammed quantum execution, where isolating independent workloads is essential.

\subsection{Limitations and Future Work}

While \system demonstrates clear advantages in quality-aware partitioning and multi-tenant isolation, its current design exposes several tensions that point toward richer research directions.

A natural next step is to augment \system with an explicit calibration-validation stage that tests whether the provider-reported calibration data is sufficiently predictive for allocation. When the calibration signal is weak, stale, or incomplete, \system could fall back to a more conservative placement policy or re-rank candidate regions using lightweight empirical probes before enabling full multi-tenant execution.

The most immediate limitation is the \emph{region-size ceiling}.
A circuit requiring $n$ qubits can be placed in the QVM only if at least one discovered community contains $n$ or more qubits.
On devices with pronounced quality heterogeneity, the highest-quality communities tend to be small precisely because strong modularity boundaries reflect genuine physical discontinuities in gate quality.
This means that moderate-scale circuits, those well within the processor's total qubit count but exceeding every individual community, currently fall outside \system's scope.
Left unaddressed, these circuits would have to be compiled against the full device graph, forfeiting the quality and isolation guarantees that motivate virtualisation in the first place.

A natural resolution is to integrate \system with the circuit cutting and knitting framework~\cite{Tang2021,tornow2024scaling}.
Rather than treating community boundaries as hard allocation limits, the system could impose a per-fragment error budget and, when no single community satisfies it for a given circuit, automatically invoke circuit partitioning.
The resulting sub-circuits would each be mapped onto a smaller, high-quality community that \emph{does} meet the error threshold, and their joint output would be reconstructed via classical knitting.
This effectively turns the limitation of community size into an opportunity: circuits that would otherwise produce noise-dominated results on a loosely compiled full device could instead execute reliably across multiple curated regions.
The approach also aligns well with the gate virtualisation paradigm introduced for QVM scaling~\cite{tornow2024scaling}, and extends it by grounding the partitioning decision in empirical device quality rather than circuit size alone.
Exploring the interplay between community structure, cutting overhead, and achievable output quality in this setting is, in our view, a particularly compelling direction.

A second tension arises between quality awareness and placement flexibility.
Our experiments show that for medium-size workloads (roughly 5--8 qubits), the unconstrained baseline occasionally finds placements that outperform \system, because it can search across the entire device coupling graph rather than within a single community.
This suggests that rigid community boundaries may sometimes exclude high-quality placements that happen to straddle two regions.
Hybrid strategies, for instance, allowing controlled boundary relaxation when inter-region coupler quality exceeds a dynamic threshold, or maintaining a secondary pool of ``soft-boundary'' regions for circuits that are borderline in size, could recover some of this flexibility without abandoning the quality-aware allocation model.

Third, \system's region discovery currently operates offline against calibration snapshots, and therefore cannot track intra-day drift or short-timescale fluctuations.
While calibration data provides a stable baseline, real devices exhibit continuous parameter drift between calibration cycles.
A practical enhancement would be to interleave lightweight benchmarking circuits, including randomised benchmarking fragments or mirror circuits, with production workloads, using their outcomes as execution-driven quality signals.
These in-situ measurements could feed into an incremental update mechanism that adjusts region scores and, when drift is severe enough, triggers partial re-discovery of community boundaries.
Such a scheme would complement static calibration with real-time, workload-relevant feedback, narrowing the gap between the assumed and actual device state.

Finally, contemporary noise models in \system capture predominantly stochastic error channels.
Real hardware, however, exhibits coherent error components including systematic over-rotations, residual $ZZ$ coupling, and calibration-dependent phase errors, which accumulate constructively rather than averaging out.
Incorporating mitigation techniques such as dynamical decoupling sequences, calibrated echo insertions, or compiler-level coherent error compensation into the \system pipeline would address an error regime that quality-weighted partitioning alone does not reach.

\subsection{Implications for Other Hardware Architectures}
The evaluations in this paper empirically validate \system on two distinct superconducting architectures: IBM's heavy-hex processors and Rigetti's square-lattice Ankaa-3 backend. More broadly, the underlying graph formulation depends only on connectivity and calibration-quality metadata rather than on any topology-specific template. The following observations, therefore, describe how the same abstraction is likely to behave on other hardware families; they are architectural implications rather than additional empirical claims.

\begin{itemize}[nosep,leftmargin=*]
\item \textbf{Dense 2D lattices (e.g., Google Sycamore):} On grid-like superconducting layouts with mostly local degree-4 connectivity, the behaviour is likely to resemble our observations on Rigetti more closely than those on IBM heavy-hex devices. In such settings, \system would be expected to trade reduced routing overhead against potentially stronger local crosstalk, with benefits concentrated in routing-sensitive workloads.

\item \textbf{Trapped-ion processors (e.g., IonQ, Quantinuum):} In architectures with effectively all-to-all or very dense logical connectivity, routing pressure is much weaker than on nearest-neighbour superconducting hardware. There, the role of \system would shift from discovering sparse topological subgraphs to stratifying execution regions based on calibration quality, laser-addressing fidelity, and time-varying qubit reliability.

\item \textbf{Neutral-atom arrays (e.g., QuEra):} For platforms with partially reconfigurable interaction geometry, the same graph-based abstraction could in principle inform both virtual-region selection and hardware layout decisions before execution. In this case, \system would no longer operate purely as a passive partitioner, but as a guide for choosing a favourable interaction structure for a given workload.
\end{itemize}

In this sense, the practical prerequisite for deploying \system is not a specific topology but rather sufficiently exposed backend telemetry that describes connectivity and execution quality. Where providers expose such calibration information at usable granularity, the same graph-driven virtualisation framework can be instantiated without redesigning the core algorithm around a new hardware family.

\section{Related Work}

\subsection{Quantum Virtual Machines and Multiprogramming} 
The concept of quantum resource virtualisation has evolved through several generations. Das et al.~\cite{das2019case} first demonstrated the feasibility of quantum multiprogramming, establishing core requirements: disjoint allocation, independent compilation, and crosstalk mitigation. This foundational work showed near-linear throughput improvements through careful qubit partitioning.

Liu and Dou~\cite{liu2021qucloud} proposed QuCloud, which applies community detection to cloud-oriented multiprogramming. Their approach also uses community detection for cloud-oriented multiprogramming, but it operates on unweighted graphs and is not designed to rank candidate regions by calibration-derived execution quality. Relative to that formulation, \system uses calibration-weighted graphs, together with an explicit region-scoring stage, to differentiate candidate regions by their expected suitability.

HyperQ~\cite{Tao2025} pioneered the QVM abstraction, introducing fixed-size virtual machines (I-shaped 7-qubit regions) that abstract physical hardware topology. Programs target QVMs rather than physical qubits, enabling hardware-independent development. HyperQ's topology-aware design achieves high coverage on heavy-hex processors but requires manual region engineering for each topology family. \system shares HyperQ's QVM goal but dynamically discovers regions, enabling topology independence and quality differentiation.

Niu et al.~\cite{Niu2023enabling} developed formal capacity models for multiprogramming, providing theoretical foundations for partition sizing. Their analysis of interference bounds complements our empirical crosstalk isolation results.

\subsection{Quantum Resource Theory and Programming}
Ying's foundational work~\cite{ying2016foundations} on quantum programming semantics provides formal foundations for reasoning about quantum resource management. Recent developments in quantum resource theories~\cite{chitambar2019} characterise quantum resources in terms of their inter-convertibility, though practical systems must also address physical device constraints. Our quality-weighted approach bridges theory and practice: the graph model encodes physical error characteristics while community detection provides principled optimisation.

\subsection{Gate virtualisation and Circuit Cutting} 
Tornow et al.~\cite{tornow2024scaling} introduced gate virtualisation for scaling quantum computations beyond hardware limits, decomposing circuits into fragments with classical communication overhead. Tang et al.~\cite{Tang2021} developed CutQC for efficient circuit cutting with reduced sampling overhead. These approaches address a complementary problem: enabling large circuits to exceed hardware size, while \system enables multiple small circuits to share hardware. The techniques are orthogonal and can be combined: large circuits can be partitioned into fragments, each assigned to a different QVM region.

\subsection{Noise-Aware Compilation} 
Murali et al.~\cite{Murali2019} proposed noise-adaptive compilation that routes circuits through low-error hardware paths. Tannu and Qureshi~\cite{Tannu2019} demonstrated variability-aware mapping for NISQ devices. These techniques optimise circuit placement \emph{within} a fixed region; \system optimises the \emph{choice} of region itself. The approaches are complementary: \system selects a high-quality region, then noise-aware compilation optimises routing within it.

\subsection{Classical virtualisation and Cloud Computing} 
Our design draws inspiration from classical hypervisor architectures~\cite{Barham2003}, particularly the separation of resource management (offline discovery) from runtime scheduling (online allocation). The high-cohesion/low-coupling principle guiding QVM boundary placement reflects decades of systems research on module design. Multi-tenant isolation mechanisms in classical clouds, including memory protection, resource quotas, and performance isolation, inform our crosstalk-suppression approach, although quantum interference requires different technical solutions.

\section{Conclusion}
We presented \system, a dynamic topology-agnostic quantum virtual machine that discovers QVM regions via quality-weighted community detection. Unlike prior QVM systems, which require manual engineering, \system derives regions directly from calibration-weighted device graphs. This topology-agnostic formulation dynamically adapts to calibration changes, producing quality differentiation regions for allocation. 

We validated \system on IBM heavy-hex and Rigetti square-lattice backends.  and discuss how the same graph formulation can be extended to other architectures that expose connectivity and calibration data. These experiments show \system is most valuable on heterogeneous hardware, where quality-aware region selection raises the performance floor and reduces worst-case failures. Batch experiments further show stable multi-tenant quality enabling a substantial reduction in job count without systematic loss of output-similarity.

As quantum processors scale and hardware architectures diversify, topology-agnostic QVM mechanisms provided by \system will become increasingly valuable for practical multi-tenant quantum cloud services.

\section*{Acknowledgements}
This work was supported by resources provided by the Pawsey Supercomputing Research Centre with funding from the Australian Government and the Government of Western Australia. It was carried out within the Pawsey Supercomputing Research Centre’s Quantum Supercomputing Innovation Hub, made possible by a grant from the Australian Government through the National Collaborative Research Infrastructure Strategy (NCRIS). Computational resources were provided by the Pawsey Supercomputing Research Centre’s Setonix Supercomputer (\href{https://doi.org/10.48569/18sb-8s43}{https://doi.org/10.48569/18sb-8s43}), with funding from the Australian Government and the Government of Western Australia.

\section*{Data Availability}
The code and evaluation artefacts are publicly available at \url{https://github.com/PawseySC/dynamicQVM}. 
\newpage
\bibliographystyle{ACM-Reference-Format}
\bibliography{dynq}

\newpage
\appendix

\section{Allocation Extensions Under Fragmentation and Contention}
\label{app:alloc-extensions}

\subsection{Multi-Region Composition for Large Circuits}
\label{sec:composition}

\paragraph{Motivation.}
Online allocation in Section~\ref{sec:online} selects a \emph{single} available QVM atomic region. This fails when the offline pool intentionally favours compact, high-cohesion QVM atomic regions: the chip may have enough free qubits in aggregate, yet no single free QVM atomic region can satisfy $|R|\ge n$. Multi-region composition is an online fallback that constructs a composed QVM that merges multiple \emph{available} QVM atomic regions into one \emph{connected} footprint, preserving the same isolation guarantee as subgraph-restricted compilation (Section~\ref{sec:hardware}). The purpose of multi-region composition is to close the fragmentation gap.

\paragraph{Formulation.}
Let $\mathcal{R}_\text{free}$ be the set of currently available atomic regions and let the request require $n$ qubits. We seek $\mathcal{C}\subseteq \mathcal{R}_\text{free}$ such that $V(\mathcal{C})=\bigcup_{R\in\mathcal{C}}V(R)$ satisfies:
(i) $|V(\mathcal{C})|\ge n$ (capacity),
(ii) regions are disjoint (inherited from the pool and availability),
(iii) the induced subgraph $G[V(\mathcal{C})]$ is connected (hard constraint),
and (iv) bridge-induced quality loss is explicitly penalised.

\paragraph{Bridge-aware greedy composition.}
To fulfil resource requests that exceed the capacity of a single atomic region, \system employs a multi-region composition strategy. We lift the hardware graph into a region-adjacency view, where two regions $R_i$ and $R_j$ are adjacent if there exists at least one coupler crossing their boundary. We define the \emph{bridge edge set} between them as:
\begin{equation}
  E_B(R_i,R_j)=\{(q_a,q_b)\in E:\ q_a\in V(R_i),\, q_b\in V(R_j)\}.
\end{equation}
Since community detection algorithms often use weaker couplers as natural boundaries for partitioning, these inter-region bridges represent a critical bottleneck to fidelity and must be explicitly scored.

Finding the exact optimal connected subgraph is NP-hard. Consequently, we adopt a heuristic approach detailed in Algorithm~\ref{alg:composition}. The process starts with a high-quality ``seed'' region and iteratively expands the footprint by attaching the optimal adjacent region.

\begin{algorithm}[t]
\caption{Greedy QVM Multi-Region Composition}
\label{alg:composition}
\begin{algorithmic}[1]
\small
\Require Available regions $\mathcal{R}_\text{free}$, target size $n$, hardware graph $G$
\Ensure Composed footprint $V$ (as a union of regions) or $\bot$
\State Choose seed $R_\text{seed}\in\mathcal{R}_\text{free}$ (largest-first); $\mathcal{C}\gets\{R_\text{seed}\}$; $V\gets V(R_\text{seed})$
\While{$|V|<n$}
  \State $R^\star\gets\bot$; $s^\star\gets-\infty$
  \For{$R\in\mathcal{R}_\text{free}\setminus\mathcal{C}$}
    \If{$V(R)\cap V\neq\emptyset$} \State \textbf{continue} \EndIf
    \State $E_B \gets \{(u,v)\in E:\ u\in V,\ v\in V(R)\}$ \Comment{Bridges to current footprint}
    \If{$E_B=\emptyset$} \State \textbf{continue} \EndIf
    \State $s \gets 0.4\,Q(R) + 0.4\,S_\text{conn}(G[V\cup V(R)]) + 0.2\,S_\text{bridge}(E_B)$
    \If{$s>s^\star$} \State $R^\star\gets R$; $s^\star\gets s$ \EndIf
  \EndFor
  \If{$R^\star=\bot$} \State \Return $\bot$ \EndIf
  \State $\mathcal{C}\gets\mathcal{C}\cup\{R^\star\}$; $V\gets V\cup V(R^\star)$
\EndWhile
\If{$G[V]$ is disconnected} \State \Return $\bot$ \EndIf
\State \Return $V$
\end{algorithmic}
\end{algorithm}

\noindent
The marginal score $s$ (Line~12) balances three factors: the intrinsic quality $Q(R)$, the structural integrity of the \emph{combined} footprint $S_\text{conn}$, and the interface quality $S_\text{bridge}$. The weights $(0.4, 0.4, 0.2)$ mirror the allocation policy in Section~\ref{sec:online}, where the \emph{bridge} score replaces the \emph{size} score as the scenario-specific constraint.
We define $S_\text{bridge}(E_B)$ by applying the gate quality score metric (Eq.~\ref{eq:gate_quality}) to the bridge set $E_B$:
\begin{equation}
    S_\text{bridge}(E_B)=\max\!\left(0,\ 1-100\cdot\frac{1}{|E_B|}\sum_{(i,j)\in E_B}\epsilon_{ij}\right).
\end{equation}
This formulation ensures that the expansion prioritises regions connected by high-fidelity couplers, preventing the formation of weak links in the composed QVM.

This keeps composition decisions aligned with the paper’s global quality semantics: prefer high-quality QVM atomic regions, reject merges that rely on weak boundary couplers, and avoid shapes with poor routing flexibility (low induced $S_\text{conn}$).

\paragraph{Integration, compilation, and trade-offs.}
If the greedy construction returns a connected footprint $V$, \system materialises a temporary composed region $\widehat{R}$ with $V(\widehat{R})=V$ and coupling map given by the induced subgraph $G[V]$. The allocator treats $\widehat{R}$ as a candidate region using the same size-matching and fitness logic as Section~\ref{sec:online}; on acceptance, all constituent atomic regions are marked busy and released together. Compilation remains strictly subgraph-restricted to $G[V]$ (Section~\ref{sec:hardware}). To mitigate boundary bottlenecks, \system additionally exposes bridge metadata (bridge edges/qubits and component boundaries) to guide initial layout and routing heuristics.

\emph{Example.} For a 12-qubit request where the largest free QVM atomic region has 8 qubits, composition seeds with $R_A$ ($|R_A|=8$, $Q(R_A)=0.75$). Two adjacent candidates are $R_B$ ($|R_B|=5$, $Q=0.70$, stronger bridges) and $R_C$ ($|R_C|=4$, $Q=0.80$, weaker bridges). Despite a higher $Q$ for $R_C$, the marginal score selects $R_B$ because of better bridge quality and resulting connectivity, yielding $|V|=13$ and allowing the circuit. The quality trade-off is explicit: the composed footprint is feasible, but bridge penalties quantify why it is typically weaker than the best single region.

Composition is a fallback: it can reduce quality (bridges are usually weaker), increase routing overhead (boundary chokepoints), and span a larger physical area (potentially exposing to more correlated noise). In exchange, it converts fragmentation-induced allocation failures into feasible placements while preserving the same isolation and compilation model.

\paragraph{Overhead.}
With $k$ free regions and average region size $\bar{n}$, each greedy step scans $O(k)$ candidates and evaluates bridge/connectivity statistics in $O(\bar{n})$, for $O(k^2\bar{n})$ worst-case time plus a linear-time connectivity check on $G[V]$. In practice (tens of regions), this adds millisecond-level overhead to the allocation path.

\subsection{Deferred Retry Scheduling}
\label{sec:retry}

In a multi-tenant setting, not all circuits can be allocated to every execution batch when the QVM comprises multiple regions. Temporary resource contention may leave no suitable region available even when a circuit is compatible with the hardware. Rejecting such requests immediately would unnecessarily reduce system utilisation and penalise bursty workloads. To address this, \system introduces a \emph{two-level deferred retry mechanism} that explicitly separates transient contention from fundamental infeasibility.

\paragraph{Problem setting.}
The allocation problem addressed by \system is inherently dynamic. Circuit requests arrive over time with heterogeneous qubit requirements for QVM regions, while regions are released only at batch boundaries, after previously admitted circuits have completed. As a result, allocation decisions must be made without knowledge of future arrivals and without global rescheduling across batches.

In the common case, as illustrated by Algorithm~\ref{alg:retry}, circuits are admitted in the order of their arrival whenever sufficient resources (QVM region or QVM atomic region) are available. When a circuit cannot be placed immediately due to temporary contention, it is preferable to defer the request rather than reject it outright, since feasibility may change once the current regions are freed. At the same time, the system must eventually distinguish between requests that are temporarily blocked and those that are fundamentally infeasible given the hardware limits. The scheduler must therefore balance prompt admission, controlled deferral, and decisive resolution of infeasible requests, while avoiding speculative backtracking or repeated global repartitioning that would increase latency and complexity.

\begin{algorithm}[t]
\caption{Deferred Retry Batch Scheduling}
\label{alg:retry}
\begin{algorithmic}[1]
\small
\Require Circuits $\{C_1, \ldots, C_N\}$, max batch size $B$
\Ensure Executed circuits with allocation results
\State $\text{retry\_queue} \gets \emptyset$
\State $i \gets 0$
\While{$i < N$ \textbf{or} $\text{retry\_queue} \neq \emptyset$}
    \State $\text{batch} \gets$ pop up to $B$ circuits from retry\_queue
    \State $\text{batch} \gets \text{batch} \cup$ next $(B - |\text{batch}|)$ circuits from $i$
    \State $\text{allocations} \gets \text{Allocate}(\text{batch})$
    \For{$c \in \text{batch}$ not in allocations}
        \State $\text{retry\_queue}.\text{append}(c)$ \Comment{Defer to next batch}
    \EndFor
    \State Execute(batch, allocations)
    \State Deallocate(allocations) \Comment{Release regions}
    \State $i \gets i + |\text{new circuits in batch}|$
\EndWhile
\State \Comment{Final sweep for remaining retries}
\For{$c \in \text{retry\_queue}$}
    \State Try allocate and execute $c$ individually
\EndFor
\end{algorithmic}
\end{algorithm}

At the first level, \emph{batch-level deferral} handles transient contention. Circuits that fail allocation in batch $k$ are placed into a retry queue rather than being rejected. In the subsequent batch $k+1$, deferred circuits are considered \emph{before} newly arrived requests, granting them priority access to regions released at the end of batch $k$. This priority rule ensures that circuits delayed only by temporary occupancy are admitted promptly once resources become available, preventing starvation under bursty load.

At the second level, \emph{global retry validation} distinguishes transient failure from hard infeasibility. After all regular batches have been processed, \system performs a final retry sweep in which the entire region pool is available, and no competing allocations exist. Circuits that still cannot be placed at this stage are guaranteed to exceed the capacity of every available QVM region and are therefore declared infeasible due to fundamental hardware limits rather than scheduling artefacts.

\paragraph{Convergence properties.}
Let $C_\text{fail}^{(k)}$ denote the set of circuits that fail allocation in batch $k$ under the batch-level retry policy. The two-level deferred retry mechanism exhibits the following convergence properties.

\begin{enumerate}[nosep,leftmargin=*]
\item \textbf{Bounded retry queue evolution.}
The size of the retry queue evolves according to
\begin{equation}
    |C_\text{fail}^{(k+1)}| \leq |C_\text{fail}^{(k)}| + n_\text{new},
\end{equation}
where $n_\text{new}$ denotes the number of newly arrived circuits in batch $k+1$.
In practice, the queue typically contracts rather than grows: deferred circuits are prioritised in the next batch, and all regions occupied in batch $k$ are deterministically released before batch $k+1$ begins. This prevents unbounded accumulation caused solely by transient contention.

\item \textbf{Implicit size-based filtering.}
Circuits in the retry queue are naturally filtered by size. Small and medium-sized circuits are highly likely to succeed upon retry because the offline selection process favours many compact, high-quality regions over a few large ones. In contrast, large circuits that exceed most regions' capacity will continue to fail at the batch level and be deferred to the second retry stage. This behaviour emerges without explicit size thresholds or classification.

\item \textbf{Termination via global retry validation.}
In the final retry sweep, the entire QVM atomic region pool is available, and no competing allocations exist. Any circuit that fails at this stage must require more qubits than the largest available region, and is therefore infeasible due to fundamental hardware limits rather than scheduling effects, thereby guaranteeing termination of the retry process and preventing indefinite deferral.
\end{enumerate}

Taken together, these properties ensure that the two-level retry mechanism converges efficiently for heterogeneous workloads. Circuits delayed by temporary resource contention are admitted promptly, often in the immediately subsequent batch, while genuinely infeasible requests are identified decisively in the global retry stage. As a result, deferred retry improves allocation success without introducing starvation or unbounded queue growth.

However, \system currently assumes that each submitted circuit fits within a single virtual region. Handling circuits that require partitioning across multiple regions is an important future direction. In such cases, subcircuits may ideally be scheduled within the same calibration window
to ensure consistent device characteristics across executions.
Nevertheless, if virtual regions remain stable across calibration windows, \system may still support robust execution without strict synchronisation.

\section{Full Experimental Setup and Reproducibility Details}
\label{app:full-setup}

This appendix provides the full experimental configuration underlying the results in Section~\ref{sec:eval}. We report the hardware platforms, calibration data sources, workload construction, execution procedure, baselines, and evaluation metrics in sufficient detail to support reproducibility and to clarify the scope of our claims.

\subsection{Platforms and Hardware Scope}
\label{app:platforms}

We evaluate \system on IBM Quantum processors with heavy-hex topology using both calibration-derived noise simulation and execution on production devices. In the simulation study, we consider five publicly accessible IBM backends whose calibration metadata are aggregated over the full chip:

\begin{itemize}[nosep,leftmargin=*]
    \item \textbf{IBM Pittsburgh}: mean two-qubit (ECR) gate error $\overline{\epsilon}_{2\mathrm{q}} = 0.8\%$
    \item \textbf{IBM Fez}: $\overline{\epsilon}_{2\mathrm{q}} = 1.0\%$
    \item \textbf{IBM Torino}: $\overline{\epsilon}_{2\mathrm{q}} = 1.2\%$
    \item \textbf{IBM Kingston}: $\overline{\epsilon}_{2\mathrm{q}} = 1.4\%$, with high spatial variance ($\mathrm{std} = 0.7\%$)
    \item \textbf{IBM Marrakesh}: $\overline{\epsilon}_{2\mathrm{q}} = 1.6\%$
\end{itemize}

Unless otherwise stated, these calibration-derived statistics correspond to the snapshot collected on 9 Dec.\ 2025. In all cases, the native two-qubit gate is echo cross-resonance (ECR).

For real-device validation, we use two production IBM Quantum backends:
\begin{itemize}[nosep,leftmargin=*]
    \item \textbf{IBM Kingston}, chosen for its pronounced spatial heterogeneity in gate quality
    \item \textbf{IBM Torino}, chosen as a contrasting backend with more moderate error levels and variance
\end{itemize}

The purpose of this combination is to evaluate \system under both controlled calibration-derived simulation and live-device conditions, while keeping the hardware family fixed so that observed differences are attributable primarily to quality heterogeneity rather than to unrelated architectural changes.

\subsection{Simulation and Execution Environment}
\label{app:sim-env}

Calibration-derived noise simulation is performed using Qiskit and Qiskit Aer~\cite{JavadiAbhari2024}. Noise models are constructed directly from IBM Quantum calibration metadata, using per-qubit and per-coupler quality information exposed by the backend. Real-device experiments are executed on IBM Quantum hardware without modifying the compiled circuits after allocation.

The computationally intensive components of \system, including hardware graph construction, community detection, region scoring, selection logic, and repeated evaluation across calibration snapshots and backends, are executed on the Setonix Quantum partition under the Setonix-Q pilot merit allocation scheme. This partition comprises four GH200 compute nodes~\cite{NVIDIACorporation2023,NVIDIACorporation2025}. Each node contains four NVIDIA Grace Hopper GH200 superchips, each of which integrates a 72-core Arm Neoverse~V2 CPU with 120\,GB of LPDDR memory and an NVIDIA Hopper H100 GPU with 96\,GB of HBM, all connected via a unified-memory architecture. Consequently, each node exposes approximately 855\,GB of addressable memory ($4\times120 + 4\times96$\,GB). Nodes are interconnected through dual Slingshot network interfaces and provide 3.6\,TB of local NVMe storage for temporary data. We use this platform to accelerate the offline discovery and evaluation pipeline and to make repeated multi-backend sweeps practical.

Importantly, our simulations do not attempt to emulate the full quantum state of a 133- or 156-qubit processor. Instead, once a circuit is assigned to a selected QVM region, the circuit is simulated only on the active qubits and couplers within that region, using the corresponding calibration-derived subgraph and noise parameters. This region-restricted execution matches the intended multiprogramming semantics, isolates the impact of quality-aware region formation, and remains computationally tractable.

\subsection{Workload Construction}
\label{app:workload}

We evaluate \system using 29 circuits from QASMBench~\cite{Li2023}, a standardised benchmark suite spanning several representative workload classes:
\begin{itemize}[nosep,leftmargin=*]
    \item quantum algorithms, including Grover search, Deutsch-Jozsa, and quantum walks;
    \item quantum arithmetic, including adders and linear-solver components;
    \item variational workloads, including VQE and QAOA with different ansätze;
    \item error-correction-related circuits, such as QEC encoding and syndrome detection;
    \item quantum simulation circuits, including Ising model, Trotter decomposition;
    \item entanglement-generation circuits, including Bell states, cat states, W states, GHZ states; and
    \item communication-style circuits, including teleportation and Simon's programs.
\end{itemize}

Circuit widths range from 2 to 10 qubits. After transpilation, circuit depths range from approximately 10 gates to more than 500 gates, thereby covering both shallow and moderately deep NISQ-style workloads. We selected this benchmark set to ensure the evaluation includes circuits with varying routing pressure, sensitivity to gate-error accumulation, and susceptibility to spatial heterogeneity.

\subsection{Evaluation Procedure}
\label{app:procedure}

For each backend, we first execute \system's offline discovery phase to derive QVM atomic regions from the backend-specific quality-weighted hardware graph. Circuits are then assigned to regions using the best-fit policy described in Section~\ref{sec:design}. After placement, each circuit is compiled and executed on the selected region.

We compare against a baseline that represents the standard execution model exposed by current quantum cloud platforms: circuits are submitted via Qiskit's default transpilation flow, without \system's quality-aware QVM region placement. This baseline preserves the usual backend-aware compilation process but does not perform community-driven partition discovery or quality-aware region selection.

Unless otherwise stated, each circuit is executed with 1024 shots. For multi-backend simulation, we use the same benchmark suite and shot count across backends to ensure consistent comparisons. For real-device evaluation, circuits are submitted to the target production backend after allocation and compilation, and the resulting counts are compared against the corresponding ideal output distributions.

For the batch-scalability experiments, workloads are grouped into batches of increasing size to study how region availability and execution quality evolve under concurrent demand. In these experiments, \system is evaluated not only by output quality but also by whether circuits can be placed onto viable regions without being forced onto low-quality or fragmented areas of the chip.

\subsection{Metrics}
\label{app:metrics}

We quantify output quality using the $L1$ distance between the measured output distribution and the ideal output distribution:
\begin{equation}
D_{L1} = \sum_x \left| p_\text{ideal}(x) - p_\text{measured}(x) \right|.
\label{eqn:dldist}
\end{equation}
This metric takes values in $[0,2]$, where smaller values indicate better agreement with the ideal distribution.

For interpretability, we also report the normalised similarity score
\begin{equation}
S = 1 - \frac{1}{2}D_{L1},
\end{equation}
where $S=1$ indicates a perfect match and $S=0$ indicates maximal disagreement. Throughout the paper, we refer to this quantity as the \emph{output-similarity score} to distinguish it from gate fidelity and from state fidelity in the strict quantum-information sense.

In addition to mean values, we report medians, lower-percentile statistics where relevant, relative improvements over the baseline, and win/loss counts at the circuit level. These complementary views allow us to distinguish average-case improvements from robustness improvements, and to identify whether gains are concentrated in only a few benchmarks or are broadly distributed across the suite.

\subsection{Reproducibility Scope and Practical Notes}
\label{app:repro}

We intend to make the evaluation reproducible in terms of methodology, workload construction, placement procedure, and metric computation. Exact numeric replication on live quantum hardware is not guaranteed over time because calibration metadata, disabled couplers, and device availability evolve across dates. This temporal variation is intrinsic to NISQ hardware and is one reason why \system explicitly relies on calibration-aware discovery rather than on static, topology-only partition templates.

The simulation results are therefore reproducible only relative to the calibration snapshots used to construct the corresponding backend noise models. The real-device results should be interpreted as date-specific measurements on production systems.

\section{Circuit-Level and Distributional Analysis}
\label{app:circuit}
This appendix complements the aggregate results in Section~\ref{sec:multibackend} by examining the distributional structure of \system's gains at the circuit level. Rather than asking only whether \system improves mean output similarity on average, we analyse how these improvements are distributed across hardware backends, across circuit families, and across individual execution outcomes. This finer-grained view helps clarify three questions: whether the observed gains are broadly distributed or driven by a small number of outliers, which workload regimes benefit most from quality-aware placement, and why mean improvement remains positive even when the raw win rate is not dominant.

\subsection{Distribution and Consistency of Circuit-Level Improvements}

\begin{figure}[t!]
    \centering
    \begin{subfigure}[b]{0.48\textwidth}
        \centering
        \Description{Boxplot chart showing distributions of output-similarity improvement across benchmark circuits on multiple quantum backends. The chart compares the variability and consistency of the \system optimisation under different device-noise conditions.}
        \includegraphics[width=\linewidth]{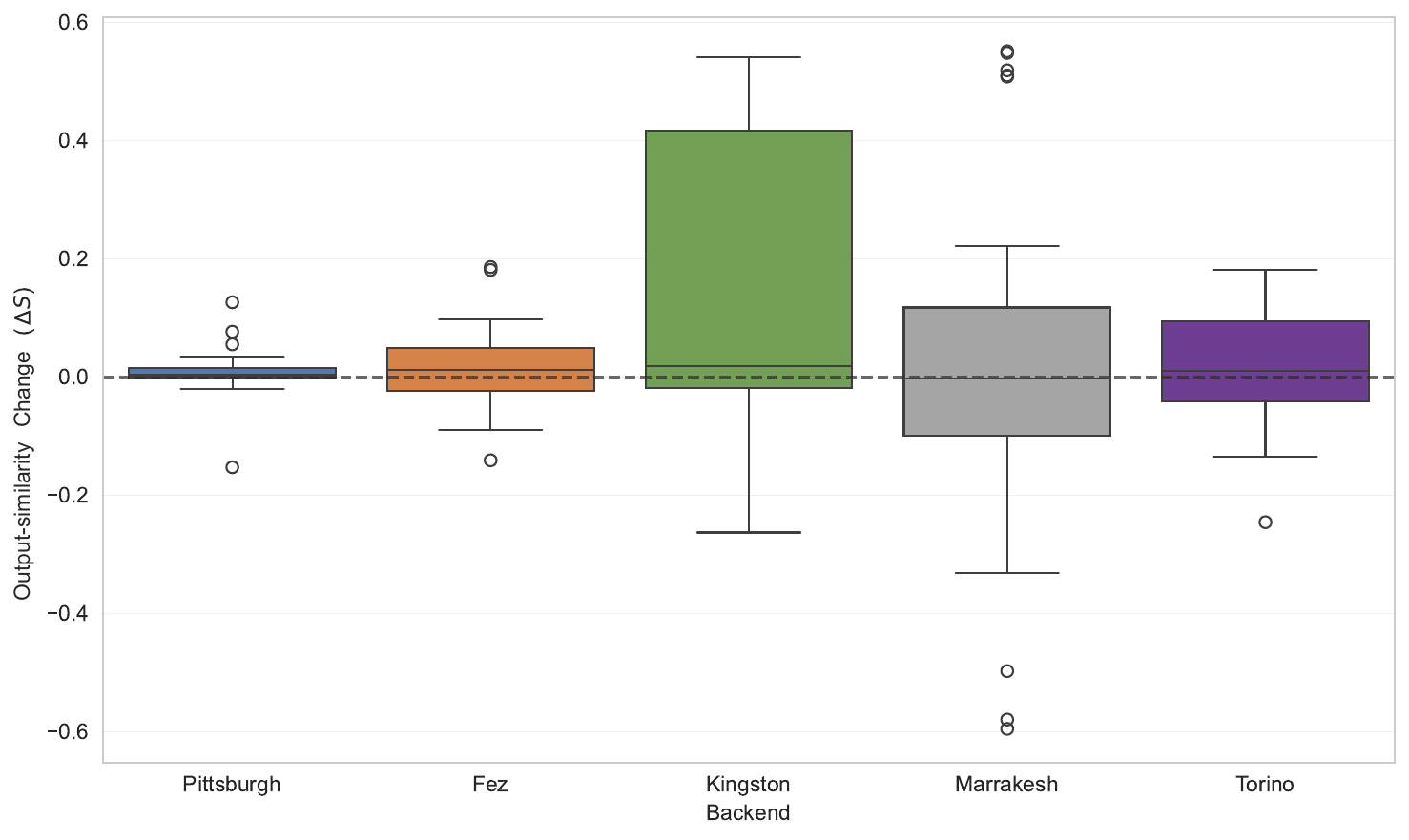}
        \caption{}
        \label{fig:circuit_analysis_dist}
    \end{subfigure}
    \hfill
    \begin{subfigure}[b]{0.48\textwidth}
        \centering
        \Description{Scatter plot showing the relationship between baseline circuit output similarity and output-similarity improvement achieved by \system. Each point represents a benchmark circuit, illustrating that larger improvements occur primarily for circuits with lower initial fidelity.}
        \includegraphics[width=\linewidth]{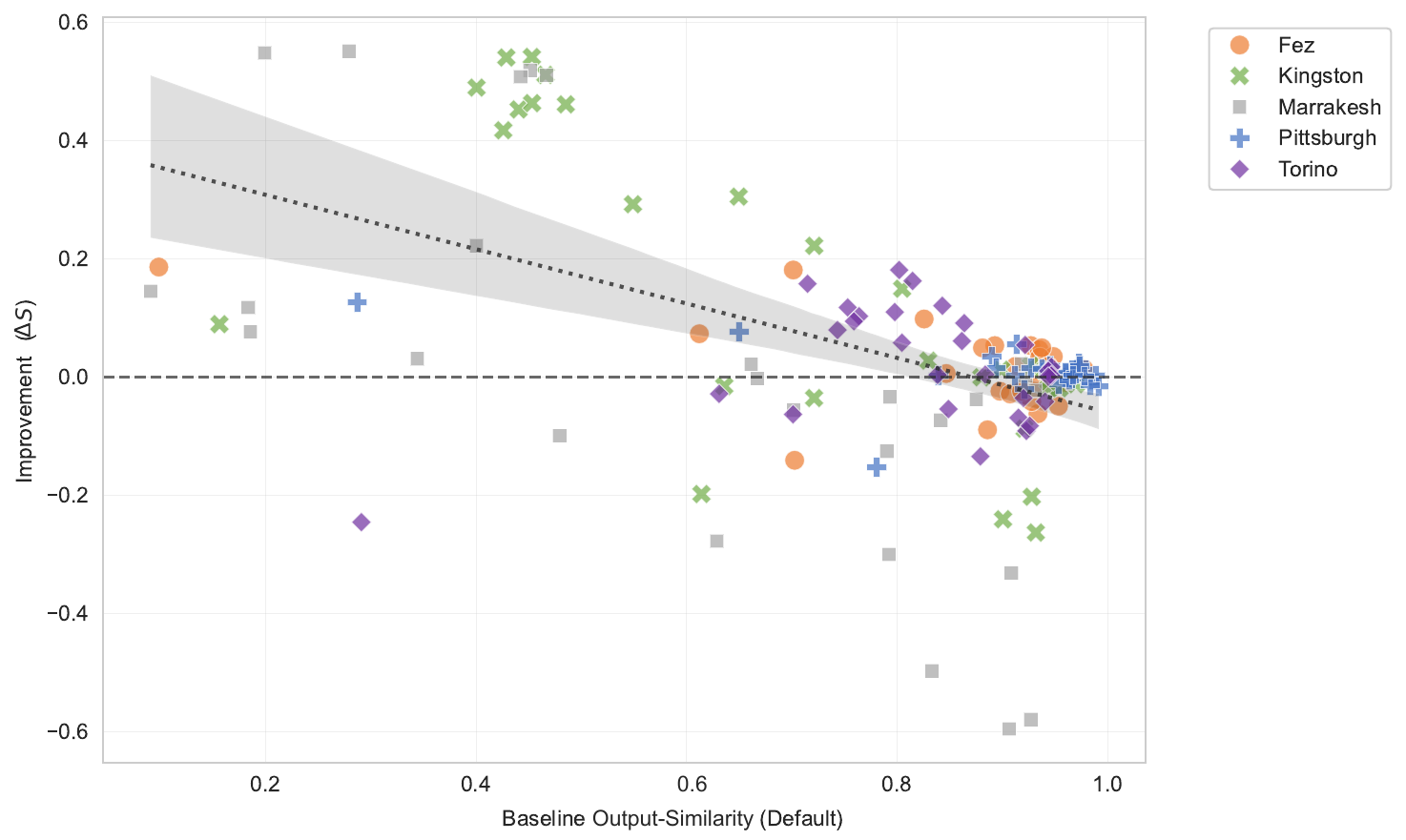}
        \caption{}
        \label{fig:circuit_analysis_scatter}
    \end{subfigure}
    \caption{\textbf{Circuit-level performance analysis.} (a) Distribution of output-similarity improvement ($\Delta S$). \system shows high consistency on stable devices (Pittsburgh, Fez) while unlocking significant ``high-reward'' potential on noisy ones (Kingston). (b) Correlation between baseline output similarity and optimisation efficacy. The strong negative trend ($r \approx -0.53$) indicates that \system primarily aids low-fidelity circuits ($<0.6$) while having a neutral impact on high-fidelity ones ($>0.9$).}
    \label{fig:circuit_analysis}
\end{figure}

While aggregate metrics establish the overall benefit, a granular analysis of circuit-level improvements reveals the specific conditions under which \system thrives. Figure~\ref{fig:circuit_analysis} provides a comprehensive view of these dynamics, contrasting backend-specific distributions with the global correlation between optimisation efficacy and baseline output similarity.

\subsubsection{Backend-Specific Performance Regimes}
As shown in Figure~\ref{fig:circuit_analysis}(a), the distribution of output-similarity gains varies significantly by backend, highlighting distinct reliability profiles. On high-fidelity backends like \textbf{Pittsburgh and Fez}, \system demonstrates \textit{high consistency}. The interquartile range is narrow and centred slightly above zero, with nearly all circuits on Pittsburgh falling within a tight band ($0$ to $+0.01$). This indicates that the compiler introduces negligible overhead on already well-optimised executions.

In contrast, \textbf{Kingston} represents a \textit{high-variance, high-reward} regime. The distribution features a broad positive tail, with nearly 45\% of circuits gaining more than 10 percentage points in output similarity. Extreme outliers, such as \texttt{basis\_test\_n4} ($+0.49$), demonstrate the system's capacity to salvage executions that would otherwise fail. However, on highly volatile backends like \textbf{Marrakesh and Torino}, the results highlight a risk–reward trade-off. While some circuits achieve massive gains (e.g., greater than $50\%$ for \texttt{grover\_n2}), others suffer degradations due to unfavourable interactions between error patterns and dynamic remapping.

\subsubsection{The Role of Baseline Output Similarity}
The variance observed across backends is primarily attributable to the initial quality of the circuit mappings. Aggregating all execution traces in Figure~\ref{fig:circuit_analysis}(b) reveals a clear inverse relationship (Pearson $r \approx -0.53$) between the baseline output similarity and the improvement provided by \system.

Circuits with low baseline output similarity ($<0.6$) consistently achieve the largest gains, confirming that dynamic optimisation effectively mitigates the effects of poor initial placements dominated by noise. Conversely, as baseline output similarity approaches the effective hardware limit ($>0.9$), the marginal benefit diminishes, stabilising near zero. In this regime, additional optimisation offers a limited advantage and may occasionally introduce minor regressions.

Ultimately, these results suggest that \system acts as a robust \textit{performance floor}: it delivers transformative improvements on heterogeneous, noise-stratified hardware while remaining safe and non-disruptive on high-fidelity devices.

\subsection{Per-Circuit Analysis}

\begin{figure}[t]
  \centering
  \Description{Bar chart showing the average output-similarity difference between \system and the default compiler for multiple benchmark quantum circuits, aggregated across simulated and real hardware backends. Positive bars indicate higher mean output-similarity achieved by \system.}
  \includegraphics[width=0.95\linewidth]{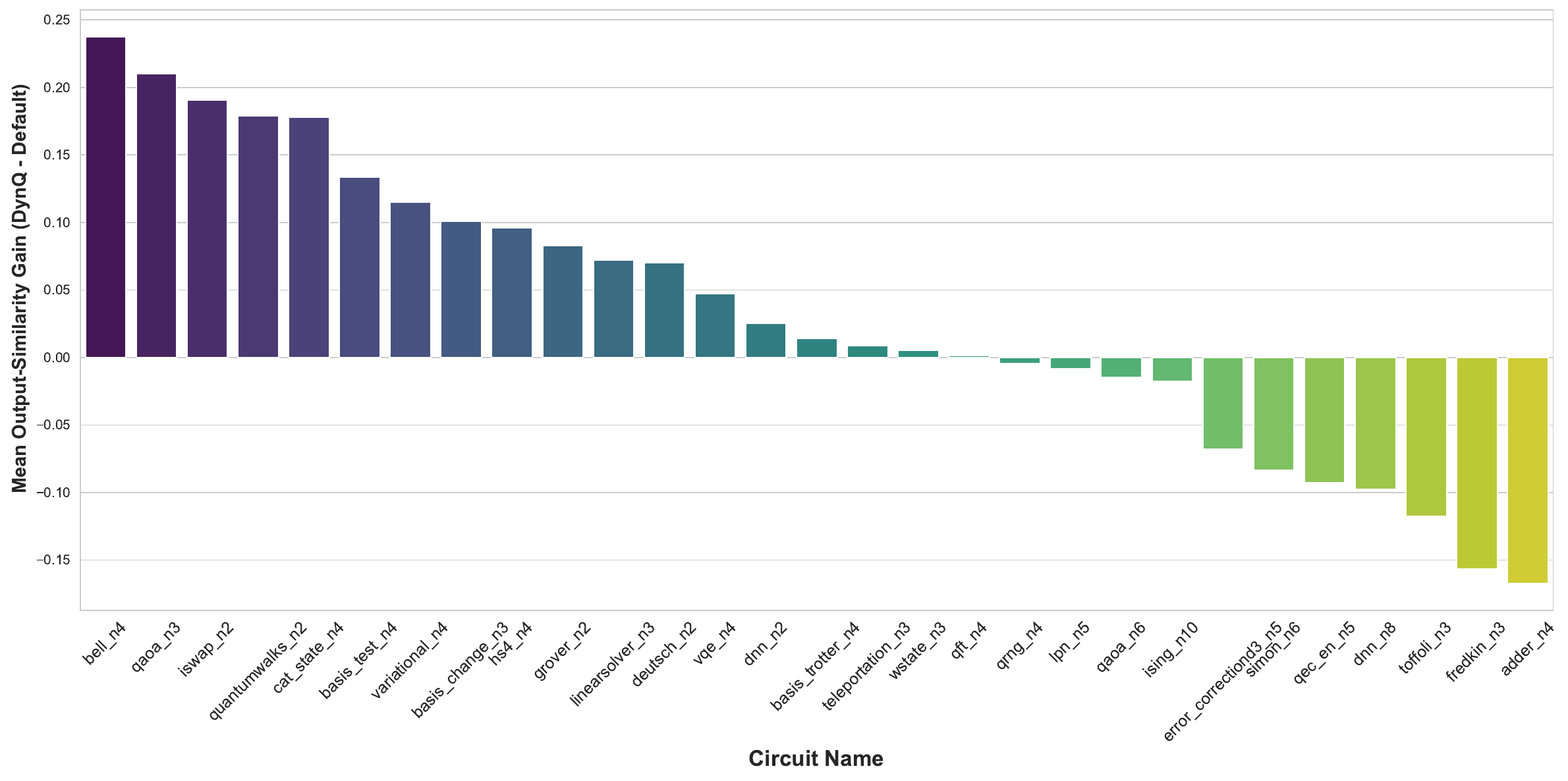}  
  \caption{\textbf{Average output-similarity improvement per benchmark circuit achieved by \system over the default compiler.} 
  The bar chart displays the mean output-similarity gain ($F_{\text{DynQ}} - F_{\text{Default}}$) for each circuit, averaged across all seven tested environments (including 5 simulated backends and 2 real quantum processors). 
  The circuits are ordered by magnitude of improvement. Positive values indicate a net performance benefit from using \system. 
  The significant gains on circuits such as \texttt{bell\_n4} highlight \system's effectiveness in optimising topology mapping and mitigating hardware errors.}  
  \label{fig:circuit_avg_improvement}
\end{figure}

Table~\ref{tab:percircuit} and Figure~\ref{fig:circuit_avg_improvement} show detailed results for selected circuits on Kingston (simulation), illustrating how different circuit characteristics respond to quality-aware placement.

\begin{table}[t]
\centering
\caption{Per-circuit analysis on Kingston (simulation)}
\label{tab:percircuit}
\small
\begin{tabular}{lcccr}
\toprule
\textbf{Circuit} & \textbf{Qubits} & \textbf{Baseline} & \textbf{\system} & \textbf{$\Delta$} \\
\midrule
\multicolumn{5}{l}{\textit{Large improvements (baseline had very low output-similarity):}} \\
quantumwalks\_n2 & 2 & 45.3\% & 99.4\% & +54.1\% \\
cat\_state\_n4 & 4 & 42.9\% & 96.9\% & +54.0\% \\
grover\_n2 & 2 & 46.6\% & 97.7\% & +51.1\% \\
iswap\_n2 & 2 & 45.3\% & 91.6\% & +46.3\% \\
linearsolver\_n3 & 3 & 48.5\% & 94.6\% & +46.1\% \\
hs4\_n4 & 4 & 44.0\% & 89.3\% & +45.3\% \\
\midrule
\multicolumn{5}{l}{\textit{Moderate improvements:}} \\
deutsch\_n2 & 2 & 64.9\% & 95.4\% & +30.5\% \\
variational\_n4 & 4 & 54.9\% & 84.1\% & +29.2\% \\
vqe\_n4 & 4 & 72.1\% & 94.3\% & +22.2\% \\
\midrule
\multicolumn{5}{l}{\textit{Limited improvement (already high or very deep):}} \\
basis\_trotter\_n4 & 4 & 15.6\% & 24.5\% & +8.9\% \\
lpn\_n5 & 5 & 97.0\% & 95.7\% & -1.3\% \\
\midrule
\multicolumn{5}{l}{\textit{Regressions (larger circuits, limited regions):}} \\
ising\_n10 & 10 & 61.4\% & 41.6\% & -19.8\% \\
simon\_n6 & 6 & 90.0\% & 65.9\% & -24.1\% \\
\bottomrule
\end{tabular}
\end{table}

\textbf{Pattern 1: Largest gains arise from avoiding severely degraded hardware regions.}
The largest improvements are observed for circuits whose baseline output similarity falls below 50\%, a regime only marginally better than random output. These cases correspond to baseline placements that route circuits through particularly noisy qubits or couplers. By explicitly selecting regions with consistently higher hardware quality, \system relocates these circuits away from degraded areas, thereby recovering near-ideal performance with output similarities in the 91--99\% range. This highlights the importance of quality-aware placement in preventing catastrophic performance loss.

\textbf{Pattern 2: Small-width circuits benefit disproportionately from quality-aware placement.}
Circuits with 2--4 qubits consistently exhibit the most significant relative improvements, often exceeding 45\%. Such circuits can be mapped entirely onto the highest-scoring regions, allowing them to exploit the best-connected and lowest-error qubits on the device. In contrast, wider circuits are constrained to larger regions that may include qubits of mixed quality, reducing the effectiveness of selective placement. This reflects a fundamental trade-off between size and quality inherent in region-based allocation. 

\textbf{Pattern 3: Extremely deep circuits show limited recoverability.}
For circuits with very large depth, quality-aware placement alone is insufficient to restore high output-similarity. The Trotter circuit, which exceeds 400 gates after transpilation, improves from 15.6\% to only 24.5\% output-similarity under \system. In this regime, accumulated gate errors dominate execution outcomes, and coherence limits are exceeded regardless of initial qubit quality. This result underscores that placement optimisations cannot compensate for excessive circuit depth on current NISQ hardware.

\textbf{Pattern 4: Performance regressions occur when circuit width exceeds high-quality region capacity.}
A small number of larger circuits, including the 10-qubit Ising simulation and the 6-qubit Simon’s algorithm, exhibit lower output-similarity under \system than under the baseline. In Kingston, the largest high-quality regions contain at most eight qubits, forcing these circuits to be allocated to larger regions with lower average quality. In contrast, the unconstrained baseline transpiler occasionally identifies favourable qubit subsets by chance across the full chip. This behaviour exposes a limitation of region-based compilation: when the circuit width exceeds the available high-quality regions, constraining placement can reduce flexibility and lead to suboptimal outcomes.

\subsection{Win/Loss Analysis}
To complement aggregate output-similarity statistics, we report a win/loss analysis that, for each circuit, indicates whether \system achieves higher output-similarity than the baseline, using a 1\% tolerance threshold to exclude changes attributable to measurement noise. Table~\ref{tab:winloss} summarises the resulting win/loss ratios across simulated backends and real hardware.

\begin{figure}[t]
  \centering
  \Description{Scatter plot comparing circuit output-similarities produced by the \system and default compilers across multiple quantum hardware backends. Each point represents a benchmark circuit, and a diagonal reference line indicates equal performance between the two methods.}
  \includegraphics[width=0.8\linewidth]{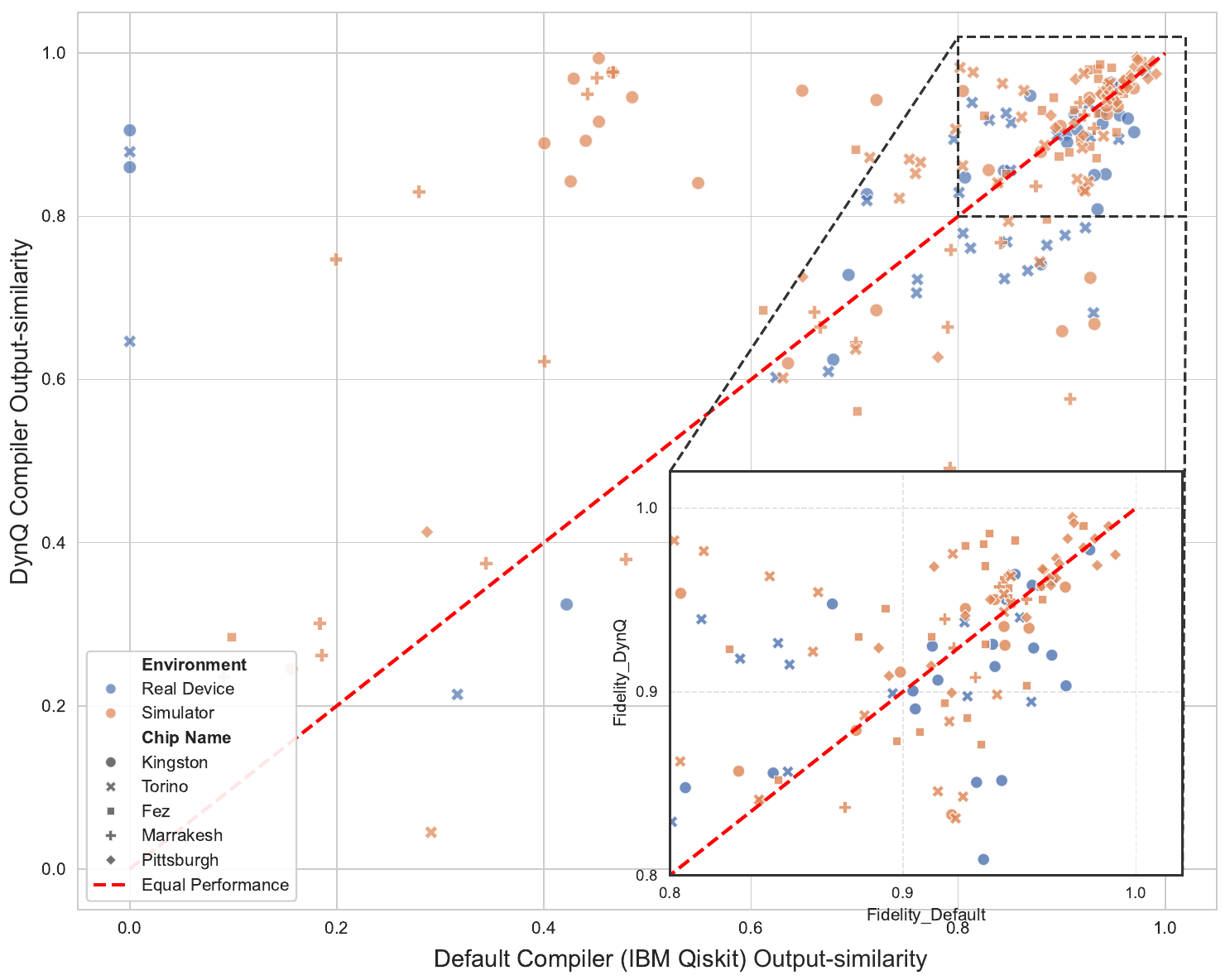}
  \caption{\textbf{Scatter plot comparison of circuit output-similarities between the default and \system compilers.} 
  The figure aggregates results from five simulated backends and two real devices.
  The diagonal line marks the break-even point. 
  The distribution reveals that \system's improvements are non-uniform: it provides substantial boosts under challenging cases (points far above the diagonal). It recovers failed executions (points on the far left), while maintaining competitive performance in the high-fidelity regime (zoomed inset). 
  This visual trend explains why the mean output-similarity gain is positive despite the mixed win rate shown in Table~\ref{tab:winloss}.}
  \label{fig:head_to_head}
\end{figure}

Figure~\ref{fig:head_to_head} provides a circuit-level, head-to-head view of output-similarity outcomes, clarifying the distributional structure underlying these aggregate statistics. While wins and losses are mixed in the high-fidelity regime, \system delivers disproportionately large gains on challenging circuits, recovers executions that fail under the default compiler, and maintains competitive performance when baseline output similarity is already high. This asymmetry explains why the mean output-similarity improvement remains positive despite a non-dominant win rate.

\begin{table}[t]
\centering
\caption{Win/loss analysis across all experiments}
\label{tab:winloss}
\small
\begin{tabular}{lcccl}
\toprule
\textbf{Setting} & \textbf{Wins} & \textbf{Losses} & \textbf{Ties} & \textbf{Win Rate} \\
\midrule
Simulation (5 backends) & 70 & 46 & 29 & 48.3\% \\
Real devices (2 backends) & 19 & 29 & 10 & 32.8\% \\
\midrule
Overall & 89 & 75 & 39 & 43.8\% \\
\bottomrule
\end{tabular}
\end{table}

At first glance, the overall win rate of 43.8\% may appear modest. However, this metric alone does not capture the asymmetric impact of improvements and regressions, nor the operational value of avoiding catastrophic failures. Three observations clarify why the win rate substantially understates the benefit of \system.

First, \emph{wins are systematically larger than losses}. Across all experiments, the mean output-similarity gain in winning cases is +18.2\%, whereas the mean regression in losing cases is -11.4\%. Consequently, the expected net improvement is positive even when wins do not constitute a strict majority. This asymmetry reflects \system's ability to deliver large gains when it succeeds, while typically incurring only moderate penalties when it does not.

Second, \emph{losses are highly structured rather than random}. Approximately 78\% of losses occur in circuits with five or more qubits, where circuit width exceeds the size of the highest-quality discovered regions. As discussed in the per-circuit analysis, this behaviour stems from an inherent trade-off of region-based allocation: constraining placement reduces flexibility for large circuits. Importantly, this limitation is predictable and confined to a well-defined subset of workloads.

Third, \emph{win/loss statistics fail to capture high-impact availability improvements}. In four cases on real hardware, the baseline compiler produced complete execution failures (0\% output similarity), whereas \system recovered functional executions with output similarity ranging from 65\% to 91\%. These events contribute disproportionately to practical utility but are counted identically to marginal wins in the win/loss metric.

Taken together, this analysis indicates that win rate is a conservative indicator of \system's value. Although \system does not improve every circuit, it delivers larger gains than losses on average, concentrates regressions in predictable regimes, and critically prevents complete execution failures. From a system-level perspective, these properties are more relevant than raw win frequency and align with the goals of robust, high-utilisation quantum cloud operation.

\section{Computational Overhead and Sensitivity Details}
\label{app:overhead}

This appendix provides additional detail on the computational cost of \system and clarifies the method's practical sensitivity to calibration-driven hardware variation. In the main text, we report only the headline conclusion that offline discovery is fast enough to be re-executed at calibration-update timescales while online allocation remains negligible. Here, we provide the detailed timing breakdown underlying that conclusion and summarise several practical sensitivity considerations that follow directly from the system design.

\subsection{Detailed Timing Breakdown}
\label{app:overhead-breakdown}

Table~\ref{tab:overhead} summarises the computational overhead of \system. Offline discovery runs once per calibration cycle and completes in 0.81\,s on IBM Kingston (156 qubits). Within this discovery phase, graph construction requires 12\,ms, community detection 89\,ms, region scoring 45\,ms, and region selection 3\,ms. The remaining time is attributable to the end-to-end orchestration of the discovery pipeline.

\begin{table}[t]
\centering
\caption{Computational overhead}
\label{tab:overhead}
\small
\begin{tabular}{lcc}
\toprule
\textbf{Operation} & \textbf{Time} & \textbf{Frequency} \\
\midrule
Offline discovery & 0.81 s & Once per calibration \\
Graph construction & 12 ms & Per discovery \\
Community detection & 89 ms & Per discovery \\
Region scoring & 45 ms & Per discovery \\
Region selection & 3 ms & Per discovery \\
\midrule
Online allocation & 0.08 ms & Per circuit \\
Transpilation (cache miss) & 0.05--30 s & First circuit of pattern \\
Transpilation (cache hit) & 0.1 ms & Subsequent circuits \\
\bottomrule
\end{tabular}
\end{table}

These results show that the additional overhead introduced by \system is concentrated in the offline stage rather than in the online allocation path. This distinction is operationally important: discovery is performed once per calibration cycle, whereas allocation is performed for every incoming circuit. Because the allocation path remains lightweight, the richer structure produced by quality-aware discovery does not translate into high runtime scheduling latency.

\subsection{Relationship to Transpilation Cost}
\label{app:overhead-transpile}

The dominant cost in the end-to-end execution path remains transpilation on a cache miss, driven by Qiskit's backend optimiser and occurring regardless of whether quality-aware QVM discovery is used. In this sense, \system does not introduce a new dominant online bottleneck. Instead, the additional cost is the discovery pass performed once per calibration cycle.

For repeated circuit structures, transpilation caching further reduces this cost. As shown in Table~\ref{tab:overhead}, cache hits reduce transpilation time from 0.05--30\,s to 0.1\,ms, corresponding to a 50{,}000$\times$ speedup in the repeated-structure case. This makes the steady-state online cost of \system effectively dominated by lightweight allocation and qubit remapping rather than by repeated backend optimisation.

\subsection{Practical Sensitivity Considerations}
\label{app:overhead-sensitivity}

The behaviour of \system is intentionally sensitive to calibration-driven hardware variation. The system is designed around the premise that qubits and couplers are not interchangeable and that virtual region boundaries should evolve as hardware conditions change. Accordingly, the discovered QVM regions are recomputed whenever the calibration state changes, rather than being treated as static templates.

This sensitivity should be interpreted as a design feature rather than as instability. In the main text, we show that the discovery phase completes in under 1 second and that the resulting regions exhibit clear quality stratification on IBM Kingston. This makes recalculation at calibration-update timescales practical.

A second practical consideration is that the computational overhead of \system is less sensitive to workload size than transpilation or execution. Discovery is backend-scoped, while online allocation remains negligible at 0.08\,ms per circuit. As a result, the principal runtime sensitivity of the system lies in execution quality rather than in scheduler latency.

Finally, the largest regressions discussed in the main text arise from structured placement trade-offs rather than from computational overhead. In particular, when baseline placement already achieves near-ceiling output-similarity, dynamic remapping can sacrifice an already favourable configuration, sometimes by introducing additional routing overhead or exposing the circuit to a less benign error structure. Conversely, in degraded or heterogeneous regions, the same discovery machinery can yield large gains in output similarity. These observations reinforce the interpretation that the cost of \system is operationally modest, while its benefits depend primarily on the structure of the backend quality landscape rather than on timing overhead.

\end{document}